\documentclass[aps,prd,superscriptaddress,preprint,tightenlines,nofootinbib]{revtex4}

\usepackage{epsfig}
\usepackage{graphicx}% Include figure files
\usepackage{graphics}
\usepackage{dcolumn}% Align table columns on decimal point
\usepackage{bm}% bold math
\usepackage{afterpage}

\newcommand{\etal}{{\it et al.}}

\begin{document}
%primary authors: Liming Zhang and Sheldon Stone

\preprint{\tighten\vbox{\hbox{\hfil CLNS 08/2044}
                        \hbox{\hfil CLEO-08-26}
                      %\hbox{\hfil S. Stone}
                      %\hbox{\hfil L. Zhang}
}}
%title
\title{Measurement of  \boldmath{${\cal{B}}(D_s^+\to\ell^+\nu)$} and the Decay Constant
\boldmath{$f_{D_s^+}$} From 600 pb$^{-1}$ of $e^+e^-$ Annihilation Data Near 4170 MeV\hspace*{2mm}}

\author{J.~P.~Alexander}
\author{D.~G.~Cassel}
\author{J.~E.~Duboscq}\thanks{Deceased}
\author{R.~Ehrlich}
\author{L.~Fields}
\author{R.~S.~Galik}
\author{L.~Gibbons}
\author{R.~Gray}
\author{S.~W.~Gray}
\author{D.~L.~Hartill}
\author{B.~K.~Heltsley}
\author{D.~Hertz}
\author{J.~M.~Hunt}
\author{J.~Kandaswamy}
\author{D.~L.~Kreinick}
\author{V.~E.~Kuznetsov}
\author{J.~Ledoux}
\author{H.~Mahlke-Kr\"uger}
\author{D.~Mohapatra}
\author{J.~R.~Patterson}
\author{D.~Peterson}
\author{D.~Riley}
\author{A.~Ryd}
\author{A.~J.~Sadoff}
\author{X.~Shi}
\author{S.~Stroiney}
\author{W.~M.~Sun}
\author{T.~Wilksen}
\affiliation{Cornell University, Ithaca, New York 14853, USA}
% ...
\collaboration{CLEO Collaboration}
\noaffiliation

%\date{\today}
\date{January 9, 2009}

\begin{abstract}
We examine $e^+e^-\to D_s^-D_s^{*+}$ and $D_s^{*-}D_s^{+}$
interactions at 4170 MeV using the CLEO-c detector in order to
measure the decay constant $f_{D_s^+}$ with good precision.
Previously our measurements were substantially higher than the most
precise lattice based QCD calculation of (241$\pm$3) MeV. Here we
use the $D_s^+\to \ell^+\nu$ channel, where the $\ell^+$ designates
either a $\mu^+$ or a $\tau^+$, when the
$\tau^+\to\pi^+\overline{\nu}$. Analyzing both modes independently,
we determine ${\cal{B}}(D_s^+\to \mu^+\nu)= (0.565\pm
0.045\pm0.017)$\%, and ${\cal{B}}(D_s^+\to \tau^+\nu)= (6.42\pm
0.81\pm0.18)$\%. We also analyze them simultaneously to find an
effective value of ${\cal{B}}^{\rm eff}(D_s^+\to \mu^+\nu)= (0.591 \pm
0.037 \pm 0.018)$\% and $f_{D_s^+}=(263.3\pm 8.2 \pm 3.9)
{~\rm MeV}$. Combining with the CLEO-c value determined independently
using $D_s^+\to\tau^+\nu$, $\tau^+\to e^+\nu\bar{\nu}$ decays, we
extract $f_{D_s^+}=(259.5\pm 6.6 \pm 3.1)$ MeV.
Combining with our previous determination of
${\cal{B}}(D^+\to \mu^+\nu)$, we extract the ratio
$f_{D_s^+}/f_{D^+}=1.26\pm 0.06\pm 0.02$. No evidence is found for a CP asymmetry
between $\Gamma(D_s^+\to \mu^+\nu)$ and $\Gamma(D_s^-\to \mu^-\nu)$;
specifically the fractional difference in rates is measured to be
(4.8$\pm$6.1)\%. Finally, we find ${\cal{B}}(D_s^+\to e^+\nu) <
1.2\times 10^{-4}$ at 90\% confidence level.
\end{abstract}

\maketitle \tighten
%\tableofcontents

%\newpage

\section{Introduction}

We discuss here an improved measurement of the width of the purely
leptonic decay $D_s^+\to\ell^+\nu$, when the $\ell^+$ is either a
$\mu^+$ or a $\tau^+$, when the latter decays into a
$\pi^+\overline{\nu}$ \cite{Previous}.
In a companion article \cite{CLEO-CSP} we report an improved measurement of the
decay width for $D_s^+\to\tau^+\nu$, where $\tau^+\to e^+\nu\bar{\nu}$.

\begin{figure}[htbp]
 \vskip 0.00cm
 \centerline{ \epsfxsize=3.0in \epsffile{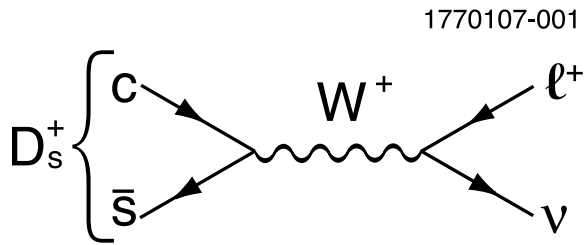} }
 \caption{The decay diagram for $D_s^+\to \ell^+\nu$.} \label{Dstomunu}
 \end{figure}

In the Standard Model (SM)
these decays are described by the annihilation of the initial
quark-antiquark pair into a virtual $W^+$ that materializes as a
$\ell^+\nu$ pair; the process is shown in Fig.~\ref{Dstomunu}.
 The decay rate is given by
\cite{Formula1}
\begin{equation}
\Gamma(D_s^+\to \ell^+\nu) = {{G_F^2}\over
8\pi}f_{D_s^+}^2m_{\ell}^2M_{D_s^+} \left(1-{m_{\ell}^2\over
M_{D_s^+}^2}\right)^2 \left|V_{cs}\right|^2~~~, \label{eq:equ_rate}
\end{equation}
where $M_{D_s^+}$ is the $D_s^+$ mass, $m_{\ell}$ is the mass of the
charged final state lepton, $G_F$ is the Fermi coupling constant,
and $|V_{cs}|$ is a Cabibbo-Kobayashi-Maskawa matrix element with a
value we take equal to $|V_{ud}|$ of 0.97418(26)
\cite{Towner-Hardy}, and $f_{D_s^+}$ is the ``decay constant," a
parameter related to the overlap of the heavy and light quark
wave-functions at zero spatial separation.

The SM decay rate then is predicted using a theoretical calculation
of the decay constant. Two calculations have been carried out using
unquenched lattice quantum-chromodynamics (LQCD). Aubin \etal~find
$f_{D_s^+}=(249\pm 3 \pm 16)$ MeV \cite{Lat:Milc}, while a more
recent calculation of Follana \etal~gives $(241\pm 3)$ MeV
\cite{Lat:Foll}. The latter calculation is more than three standard
deviations lower than the average of previous CLEO and Belle
measurements \cite{Rosner-Stone}.

Dobrescu and Kronfeld have proposed three models based on physics
beyond the SM that are consistent with known data and could possibly
explain the difference. One is a charged Higgs model and the other
two involve different manifestations of leptoquarks
\cite{Dobrescu-Kronfeld}. The recent CLEO measurement of
$f_{D^+}=(205.8\pm 8.5\pm 2.5)$ MeV, is consistent with both the
Aubin \etal~and Follana \etal~predictions, of $(201\pm 3 \pm 17)$
MeV and $(208\pm 4$) MeV, respectively \cite{ournewDp}.

It is particularly important to understand if the discrepancy in the
$D_s$ case is due to physics beyond the SM, a faulty theoretical
calculation, or to an unlikely measurement fluctuation.  We note
that precise information on the size of Cabibbo-Kobayashi-Maskawa
matrix elements extracted from $B-\overline{B}$ mixing measurements
requires theoretical input on the ``decay constants" for $B_d$
and $B_s$ mesons or their ratio, $f_{B_s}/f_{B_d}$
\cite{formula-mix}. Although the calculations in the $B$ and $D$
systems are not the same, many of the techniques used are common,
and a discrepancy in the charm system at a minimum, does not give
confidence in the theoretical predictions for the $B$ system. In
this paper we present an updated measurement of $f_{D_s^+}$ with much improved
precision.

 Akeroyd predicts
that the presence of a charged Higgs boson would suppress $f_{D_s}$
\cite{Akeroyd}. There is however the possibility, not considered by
Akeroyd, that it is the charm quark that is responsible for a NP
contribution not the $s$ quark \cite{Dobrescu-Kronfeld}. In that
case the relative change would be similar in $D^+$ and $D_s^+$
decays.

We can also measure the ratio of decay rates to different leptons,
and the SM predictions then are fixed only by well-known masses. For
example, for $\tau^+\nu$ to $\mu^+\nu$:

\begin{equation}
R\equiv \frac{\Gamma(D_s^+\to \tau^+\nu)}{\Gamma(D_s^+\to
\mu^+\nu)}= {{m_{\tau^+}^2 \left(1-{m_{\tau^+}^2\over
M_{D_s^+}^2}\right)^2}\over{m_{\mu^+}^2 \left(1-{m_{\mu^+}^2\over
M_{D_s^+}^2}\right)^2}}~~. \label{eq:tntomu}
\end{equation}
Using measured masses \cite{PDG}, this expression yields a value of
9.76 with a small error.
%After multiplying by
%${\cal{B}}(\tau^+\to\pi^+\overline{\nu}$) of (10.90$\pm$0.07)\%, the
%ratio is 1.059 for $\tau^+\nu$ with respect to $\mu^+\nu$, when
%$\tau^+\to\pi^+\overline{\nu}$.
Any deviation in $R$ from the value predicted by Eq.~(\ref{eq:tntomu})
would be a manifestation of physics beyond the SM. This could occur
if any other charged intermediate boson existed that affected the
decay rate differently than mass-squared. Then the couplings would
be different for muons and $\tau$'s. This would be a clear violation
of lepton universality \cite{Hewett}.

Most other measurements of $f_{D_s^+}$ have been hampered by a lack
of statistical precision, and relatively large systematic errors
\cite{CLEO,BEAT,ALEPH,L3,OPAL,Babar-munu}. One large systematic
error source has been the lack of knowledge of the absolute
branching fraction of the normalization channel, usually
$D_s^+\to\phi\pi^+$ \cite{stone-fpcp}. The results we report here
will not have this limitation, nor did our previous measurement
\cite{Previous}, nor did the Belle measurement \cite{Belle-munu}.

In both $\mu^+\nu$ and $\tau^+\nu$ $D_s$ decays the charged lepton must be produced with the
wrong helicity because the $D_s$ is a spin-0 particle, and the final
state consists of a naturally left-handed spin-1/2 neutrino and a
naturally right-handed spin-1/2 anti-lepton. Because the $\tau^+$
has a mass close to that of the $D_s^+$, the helicity suppression is
broken with respect to the $\mu^+$ decay, but there is an additional
large phase space suppression. Because of the helicity suppression
in $\mu^+\nu$ the radiative process $\gamma\mu^+\nu$ may have a
significant rate. Dobrescu and Kronfeld, however, estimate this
process is only 1\% of the lowest order mechanism, for photon
momenta below 300 MeV, which is relevant range for this analysis. We include this
radiative correction in what follows \cite{Dobrescu-Kronfeld, Burdman}.
(There is no correction for the $\tau^+\nu$ final state.)
The data will also be corrected for final state radiation of
the muon, as our Monte Carlo simulation incorporates this effect
\cite{photos}.

\section{Experimental Method}
\subsection{Selection of $D_s$ Candidates}
The CLEO-c detector \cite{CLEODR} is equipped to measure the momenta
and directions of charged particles, identify them using specific
ionization ($dE/dx$) and Cherenkov light (RICH) \cite{RICH}, detect
photons and determine their directions and energies.

In this study we use 600 pb$^{-1}$ of data produced in $e^+e^-$
collisions using the Cornell Electron Storage Ring (CESR) and
recorded near a center-of-mass energy ($E_{\rm CM}$) of 4.170 GeV.
At this energy the $e^+e^-$ annihilation cross-section into
$D_s^-D_s^{*+}$ + $D_s^{*-}D_s^{+}$ is approximately 1~nb, while the
cross-section  for $D_s^+D_s^-$ is about a factor of 20 smaller. In
addition, $D$ mesons are produced mostly as $D^{*}\overline{D^*}$,
with a cross-section of $\sim$5~nb, and also in
$D^*\overline{D}+D\overline{D^*}$ final states with a cross-section
of $\sim$2 nb. The $D\overline{D}$ cross-section is a relatively
small $\sim$0.2 nb \cite{poling}. There also appears to be
$D\overline{D}^*\pi$ production. The underlying light quark
``continuum" background is about 12 nb. The relatively large
cross-sections, relatively large branching fractions, and sufficient
luminosities allow us to fully reconstruct one $D_s$ as a ``tag,"
and examine the properties of the other. In this paper we designate
the tag as a $D_s^-$ and examine the leptonic decays of the $D_s^+$,
though in reality we use both charges for tags and signals. Track
requirements, particle identification, $\pi^0$, $\eta$, and $K_S^0$
selection criteria are the same as those described in
Ref.~\cite{our-fDp}, except that we now require a minimum momentum
of 700 MeV/c for a track to be identified using the RICH.

We also use several resonances that decay via the strong
interaction.
% For $\eta'\to\pi^+\pi^-\eta$ we require that the
%measured $\eta'$ mass be within $\pm$ 10 MeV of the known mass. For
%$\phi\to K^+ K^-$ we require that the reconstructed invariant mass
%be within $\pm$10 MeV of the known $\phi$ mass. For $K^{*0}\to
%K^-\pi^+$ we require that the reconstructed invariant mass be within
%$\pm$100 MeV  of the known $K^{*0}$ mass. Finally, for $\rho^-\to
%\pi^-\pi^0$ we require that the reconstructed invariant mass be
%within $\pm$150 MeV of the of the known $\rho^-$ mass.
Here we select intervals in invariant mass within $\pm$10 MeV of the
known mass for $\eta'\to\pi^+\pi^-\eta$, $\pm$20 MeV of the
known mass for $\eta'\to\rho^0\gamma$, $\pm$10 MeV for $\phi\to
K^+ K^-$, $\pm$100 MeV for $K^{*0}\to K^-\pi^+$, and $\pm$150 MeV
for $\rho^-\to \pi^-\pi^0$ or $\rho^0\to\pi^+\pi^-$.

\begin{figure}[htb]
%\centering
\includegraphics[width=74mm]{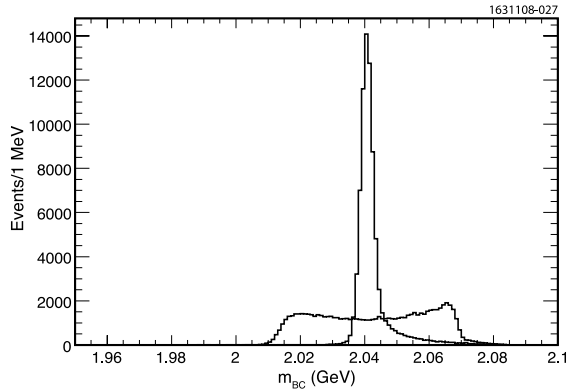}
\vspace{0.44mm}\caption{The beam constrained mass $m_{\rm BC}$ from
Monte Carlo simulation of $e^+e^-\to D_s^+D_s^{*-}$ at an $E_{\rm
CM}$ of 4170 MeV. The narrow peak is from the $D_s^+$ and the wider
one from $D_s^{*-}\to\gamma D_s^-$. (The distributions are not
centered at the $D_s^+$ or $D_s^{*+}$ masses, because the
reconstructed particles are assumed to have the energy of the
beam.)} \label{mbc}
\end{figure}

%The events we reconstruct here occur when $e^+e^-\to
%D_s^{*+}D_s^-$ or $D_s^{+}D_s^{*-}$.
We reconstruct tags from either directly produced $D_s$ mesons or
those that result from the decay of a $D_s^*$. The beam constrained
mass, $m_{\rm BC}$, is formed by using the beam energy to construct
the $D_s$ candidate mass via the formula
\begin{equation}
m_{\rm BC}=\sqrt{E_{\rm beam}^2-(\sum_i{\bf p}_{i})^2},
\end{equation}
where $i$ runs over all the final state particles. If we ignore the
photon from the $D_s^*\to\gamma D_s$ decay, and reconstruct the
$m_{\rm BC}$ distribution, we obtain the distribution from Monte
Carlo simulation shown in Fig.~\ref{mbc}. The narrow peak occurs
when the reconstructed $D_s$ does not come from the $D_s^*$ decay,
but is directly produced.

\begin{figure}[hbtp]
\centering
\includegraphics[width=6in]{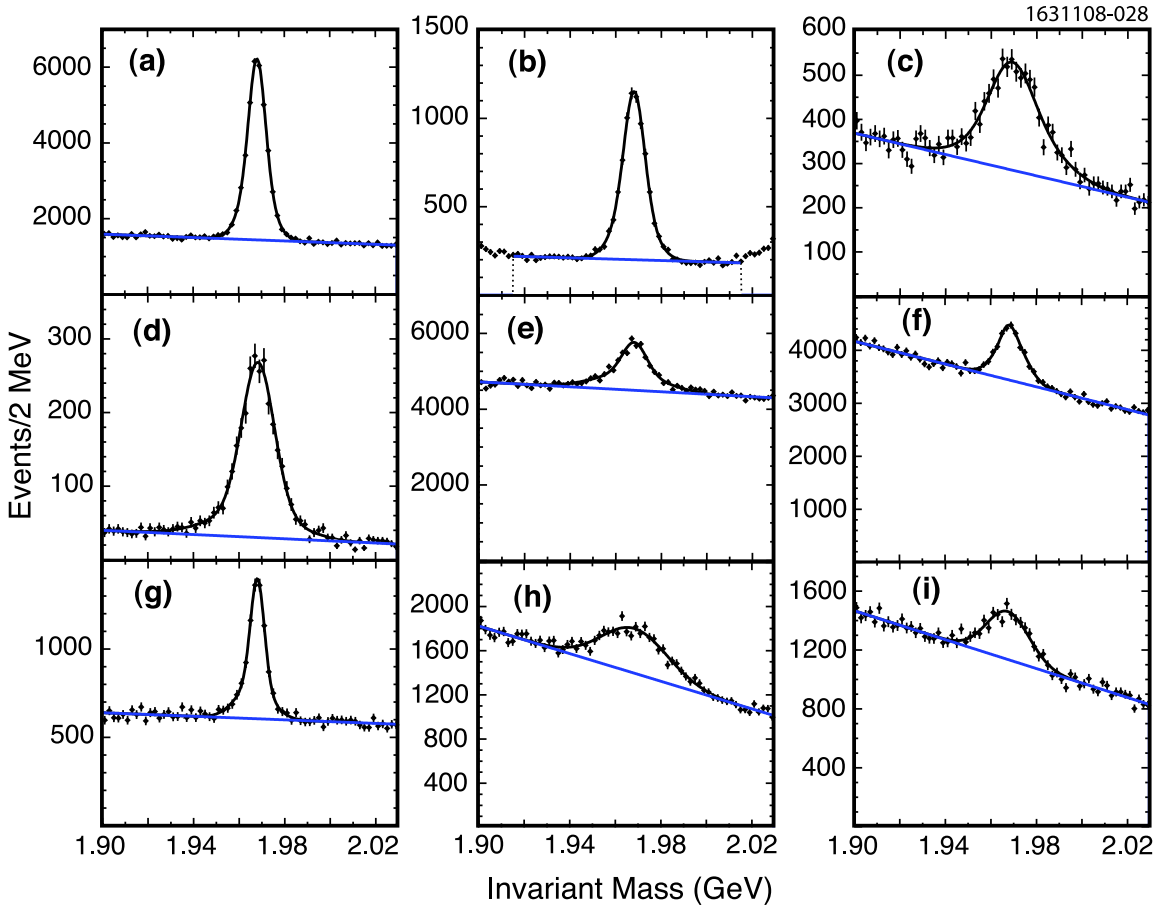}
\caption{Invariant mass of $D_s^-$ candidates in the decay modes (a)
$K^+K^-\pi^-$, (b) $K_SK^-$, (c) $\eta\pi^-$ ($\eta\to\gamma\gamma$),
(d) $\eta'\pi^-$ ($\eta'\to\pi^+\pi^-\eta$, $\eta\to\gamma\gamma$)
(e) $K^+K^-\pi^-\pi^0$ (f) $\pi^+\pi^-\pi^-$, (g) $K^{*-}K^{*0}$,
$K^{*-}\to K_S^0\pi^-$  ($K^{*0}\to K^+\pi^-$), (h) $\eta\rho^-$, and
(i) $\eta'\pi^-$ ($\eta'\to\rho^0\gamma$), after requiring the
total energy of the $D_s^-$ candidate to be consistent with the beam
energy. The curves are fits to two-Gaussian signal functions plus a
linear background. The dashed vertical lines in (b) indicate the
restricted fit region.} \label{Inv-mass}
\end{figure}

Rather than selecting events based on only $m_{\rm BC}$, we first
select an interval that accepts most of the events, $2.015 <m_{\rm
BC}<2.067$ GeV, and examine the invariant mass. Distributions from
data for the nine tag decay modes we use in this analysis are shown in
Fig.~\ref{Inv-mass}. Note that the resolution in invariant mass is
excellent, and the backgrounds not abysmally large, at least in
these modes. To determine the number of $D_s^-$ events we fit the
invariant mass distributions to the sum of two Gaussians centered at
the $D_s^-$ mass, a function we refer to as ``two-Gaussian." The
r.m.s. resolution ($\sigma$) is defined as
\begin{equation}
\sigma \equiv f_1\sigma_1+(1-f_1)\sigma_2, \label{eq:twoGauss}
\end{equation}
where $\sigma_1$ and $\sigma_2$ are the individual widths of each of
the two Gaussians and $f_1$ is the fractional area of the first
Gaussian. The number of tags in each mode is listed in
Table~\ref{tab:Ntags}. There are two changes in modes from our
previous analysis \cite{Previous}; instead of $\phi\rho^-$, we now
use $K^+K^-\pi^-\pi^0$, and we have added the $\eta'\pi^-$,
$\eta'\to\rho^0\gamma$ mode; the background here is somewhat
reduced as we apply a cut on the helicity angle, $\theta$ of
$|\cos \theta|<0.8$, since the $\rho^0$ is polarized as
$\sin^2\theta$. (Here $\theta$ is the angle of the $\pi^+$ in
the $\rho^0$ rest frame with respect to the $\rho^0$ direction
in the parent frame.)  These changes results in
an increase of 20\% more tags at the expense of more background.

\begin{table}[htb]
\begin{center}
\caption{Tagging modes and numbers of signal and background events,
within $\pm$17.5 MeV of the $D_s^-$ mass for each mode, determined
from two-Gaussian fits to the invariant mass plots, and the number
of tags in each mode including the $\gamma$ from the $D_s^*\to\gamma
D_s$ transition, within an interval $3.872 < {\rm MM}^{*2}<4.0$ GeV$^2$, as
determined from fits of the MM$^{*2}$ distributions (see text) to a
signal Crystal Ball function (see text) and two 5th order Chebychev
background polynomials.\label{tab:Ntags}}
\begin{tabular}{lcrcr}
 \hline\hline
    Mode  & \multicolumn{2}{c}{Invariant Mass}& \multicolumn{2}{c}{MM$^{*2}$}\\
    &  Signal & Background & Signal & Background \\\hline
$K^+K^-\pi^- $ & 26534$\pm$274 & 25122 &16087$\pm$373 &39563\\
$K_S K^-$ & 6383$\pm$121 & 3501 & 4215$\pm$228&6297\\
$\eta\pi^-$; $\eta\to\gamma\gamma$ & $2993\pm156$  &
5050&2005$\pm$145 &5016\\
$\eta'\pi^-$; $\eta'\to\pi^+\pi^-\eta$, $\eta\to\gamma\gamma$
& 2293$ \pm $82  &531 &1647$\pm$131 &1565 \\
$K^+K^-\pi^-\pi^0$ & 11649$ \pm $754  &78588&6441$\pm$471&89284\\
$\pi^+\pi^-\pi^-$ & 7374$ \pm $303  & 60321 & 5014$\pm$402& 43286\\
$K^{*-}K^{*0}$; $K^{*-}\to K_S^0\pi^-$, ${K}^{*0}\to K^+\pi^-$ &
4037$\pm$160& 10568&2352$\pm$176 & 12088\\
$\eta\rho^-$; $\eta\to\gamma\gamma$, $\rho^-\to \pi^-\pi^0$
& 5700$ \pm $281  &24444 & 3295$\pm$425 & 24114\\
$\eta'\pi^-$; $\eta'\to\rho^0\gamma$,
& 3551$ \pm $202  &19841 &2802$\pm$227 &17006 \\
\hline
Sum &  $70514\pm 963 $ &227966 & 43859$\pm$936&238218\\
\hline\hline
\end{tabular}
\end{center}
\end{table}

\begin{figure}[hbt]
\centering
\includegraphics[width=5in]{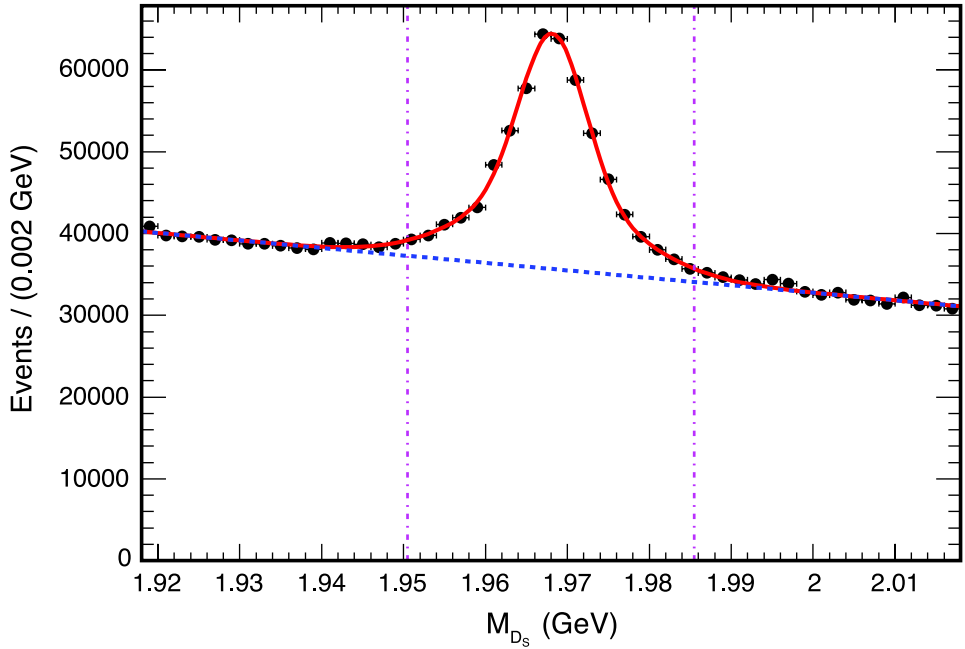}
\caption{Invariant mass of $D_s^-$ candidates
summed over all decay modes and fit to a two-Gaussian signal shape
plus a straight line for the background. The vertical dot-dashed
lines indicate the $\pm$17.5 MeV definition of the signal region.
 } \label{mass38-48_all}
\end{figure}

We list the number of signal events in each mode in
Table~\ref{tab:Ntags} by finding the number of events within
$\pm$17.5 MeV of the $D_s$ mass; here we integrate the two-Gaussian PDF
over the interval.  We also include the amount of
background in this interval. For ease of further analysis we sum all
tag modes together, as shown in Fig.~\ref{mass38-48_all}.

In this analysis we look for two types of events: (1) $e^+e^-\to
D_s^{*-}D_s^+$, and (2) $e^+e^-\to D_s^{-}D_s^{*+}$, where our
convention is that the tag is denoted by the negative charge state
and the putative signal $\mu^+\nu$ decay by the positive charge
state. Thus, we need to detect the photon from the $D_s^*$ decay.
Therefore, we look for an additional photon candidate in the event
that satisfies our shower shape requirement. Regardless of whether
or not the photon forms a $D_s^*$ with the tag, for real $D_s^*D_s$
events, the missing mass squared, MM$^{*2}$, recoiling against the
photon and the $D_s^-$ tag should peak at the $D_s^{+}$
mass-squared. We calculate
\begin{equation}
\label{eq:mmss} {\rm MM}^{*2}=\left(E_{\rm
CM}-E_{D_s}-E_{\gamma}\right)^2- \left({\bf p}_{\rm
CM}-{\bf p}_{D_s}-{\bf p}_{\gamma}\right)^2,
\end{equation}
where $E_{\rm CM}$ (${\bf p}_{\rm CM}$) is the
center-of-mass energy (momentum), $E_{D_s}$
(${\bf p}_{D_s}$) is the energy (momentum) of the fully
reconstructed $D_s^-$ tag, and $E_{\gamma}$
(${\bf p}_{\gamma}$) is the energy (momentum) of the
additional photon. In performing this calculation we use a kinematic
fit that constrains the decay products of the $D_s^-$ to the known
$D_s$ mass and conserves overall momentum and energy. All photon
candidates in the event are used, except for those that are decay
products of the $D_s^-$ tag candidate.

\begin{figure}[hbt]
\centering
\includegraphics[width=6in]{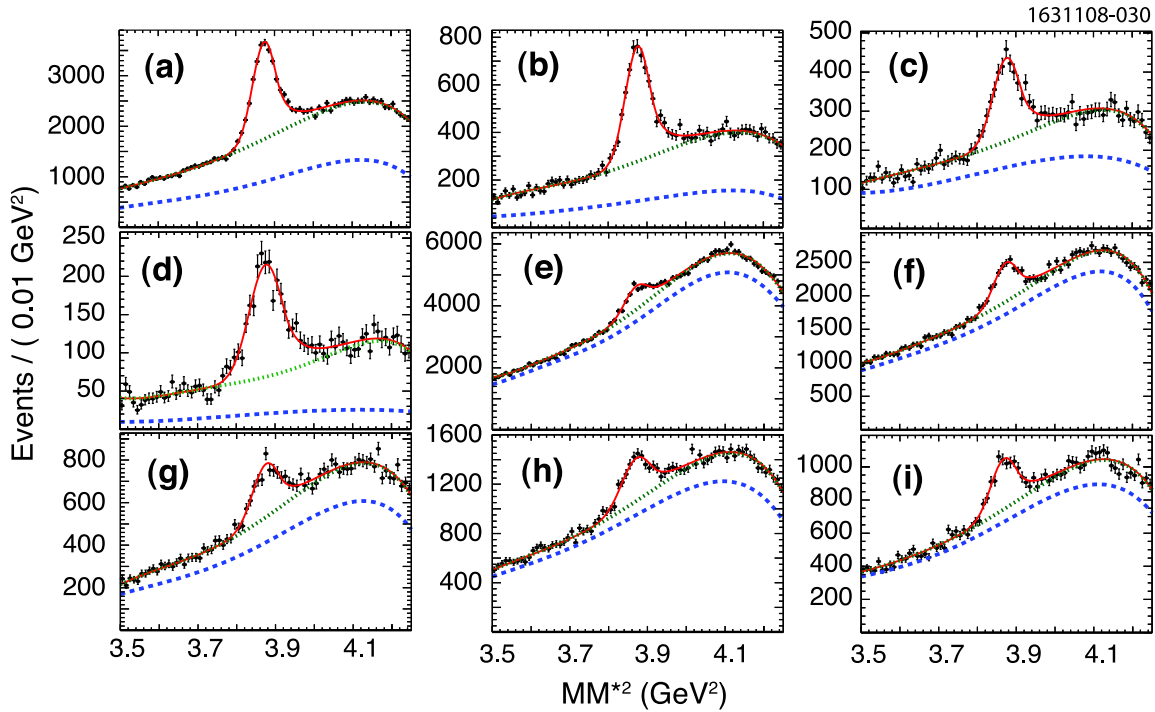}
\caption{The MM$^{*2}$ distribution from events with a photon in
addition to the $D_s^-$ tag for the modes: (a) $K^+K^-\pi^-$, (b)
$K_S^0K^-$, (c) $\eta\pi^-$, (d) $\eta'\pi^-$, (e) $K^+K^-\pi^-\pi^0$, (f)
$\pi^+\pi^-\pi^-$, (g) $K^{*-}K^{*0}$, (h) $\eta\rho^-$, and (i)
$\eta'\pi^-$, $\eta'\to\pi^+\pi^-\gamma$ . The curves are fits to
Crystal Ball functions and two 5th order Chebychev background
functions (see text).} \label{MMstar2}
\end{figure}

The MM$^{*2}$ distributions for events in the $D_s^-$ invariant mass
signal region ($\pm$17.5 MeV from the $D_s$ mass) are shown in
Fig.~\ref{MMstar2}. In order to find the number of tags used for
further analysis we preform a two-dimensional binned maximum liklihood fit of the MM$^{*2}$
distribution and the invariant mass distribution in the interval $\pm$ 60 MeV from the
$D_s$ mass and $3.5 < {\rm MM}^{*2} <4.25~GeV^2$.  This procedure is
improved by having information on the shape of the MM$^{*2}$ signal function
(often called a Probability Distribution Function, or PDF).
One possibility is to use the Monte Carlo simulation for this
purpose, but that would introduce a relatively large systematic
uncertainty. Instead, we use our relatively large sample of fully
reconstructed $D_s D_s^{*}$ events, where we use the same decay
modes listed in Table~\ref{tab:Ntags}; we find these events and then
examine the signal shape in data when one $D_s$ is ignored.

\begin{figure}[hbt]
\centering
\includegraphics[width=4in]{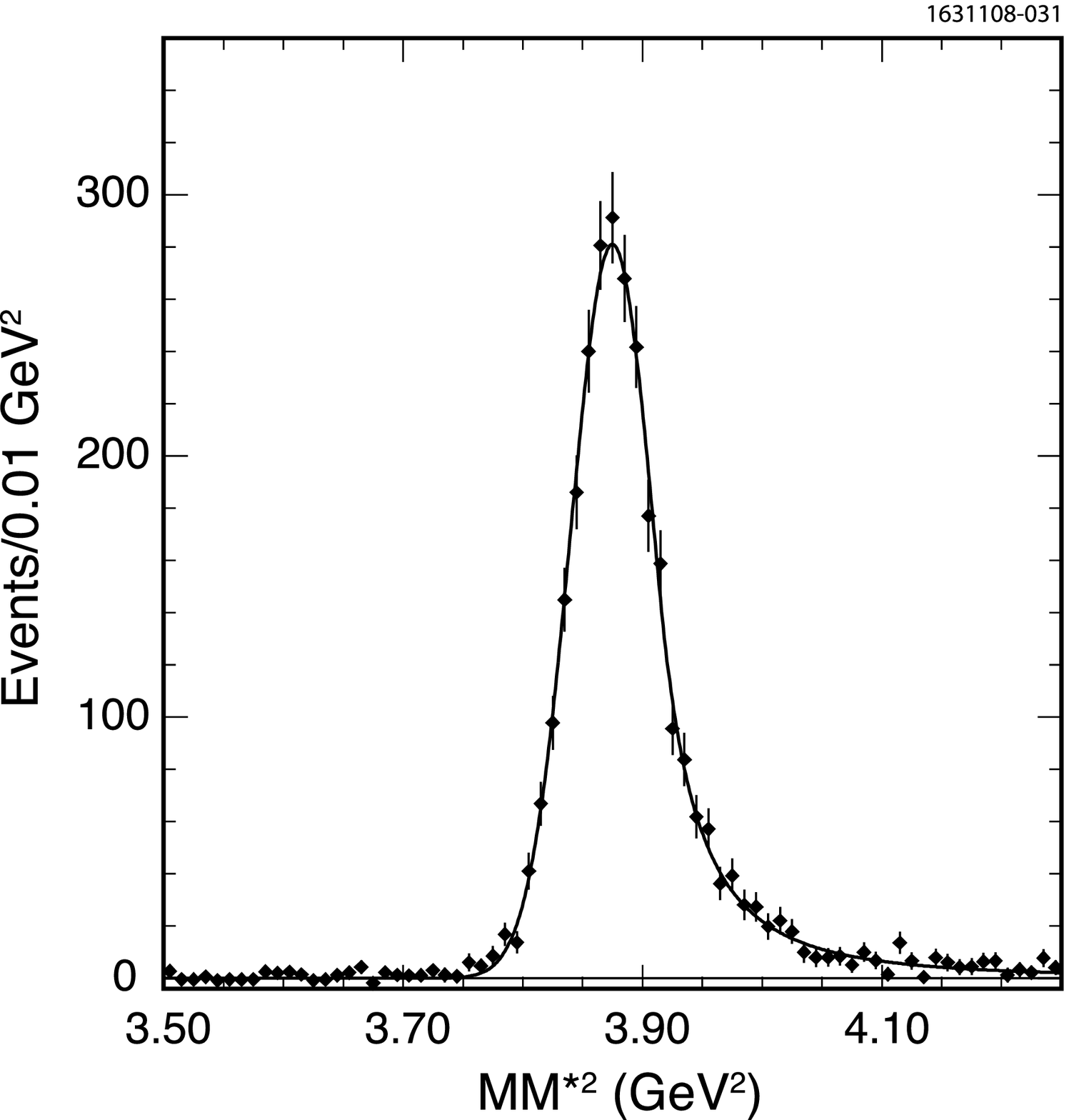}
\caption{The MM$^{*2}$ distribution from a sample of fully
reconstructed $D_s^- D_s^{*+}$ and $D_s^{*-} D_s^{+}$ events where
one $D_s$ is ignored. The curve is a fit to the Crystal Ball
function.} \label{data-Ds-MM2-DoubleTags}
\end{figure}

To remove background we subtract invariant mass sidebands. Some random photon background remains. We remove this by examining the direction of the candidate photon; it should be opposite the direction of the $D_s^+D_s^-$ system for signal. Defining the angle $\theta_{\gamma}$ to be that between the three-vector of the candidate photon and the $D_s^+D_s^-$ system, we require that $-1.0<\cos(\theta_{\gamma})<-0.95$. Furthermore we use the events for $-0.8<\cos(\theta_{\gamma})<1.0$ as background and subtract them also. The
MM$^{*2}$ distribution from this sample is shown in
Fig.~\ref{data-Ds-MM2-DoubleTags}. The signal is fit to a Crystal
Ball function \cite{CBL,taunu}. The $\sigma$ parameter, that
represents the width of the distribution, is found to be
0.035$\pm$0.001 GeV$^2$. We do expect this to vary somewhat
depending on the final state, but we do not expect the parameters
that fix the shape of the tail to change, since they depend mostly
on initial state radiation, beam energy spread, and the properties of photon detection.

The background has two components, both
described by 5th order Chebyshev polynomials; the first comes from
the background under the invariant mass peak, defined by the
sidebands, and the second is due to multiple photon combinations. In
both cases we allow the parameters to float.

 We find a
total of 43859$\pm$936 events within within the interval
$3.872 < {\rm MM}^{*2}<4.0$ GeV$^2$ and having an invariant mass within $\pm$17.5 MeV of the $D_s$ mass,
where the total number of events is the sum of the yields from the fits to
each mode as shown in Table~\ref{tab:Ntags}. An overall systematic error of
2.0\% on the number of tags is assigned by using different functions
for the description of the backgrounds. If we fix the shape of the
multiple photon combinations background polynomial to that given by
the Monte Carlo, we increase the yield by 1.1\%. If we use a 4th
order polynomial to describe the background of non-$D_s^-$ events,
we decrease the yield by 1.6\%; on the other hand using a 6th order
polynomial increases the yield by 1.7\%. Combining the results of
changing both background shapes in quadrature for the worst case, we
assign a systematic error of $\pm$2.0\% on the tag yield.

\begin{figure}[hbt]
\centering
\includegraphics[width=4.4in]{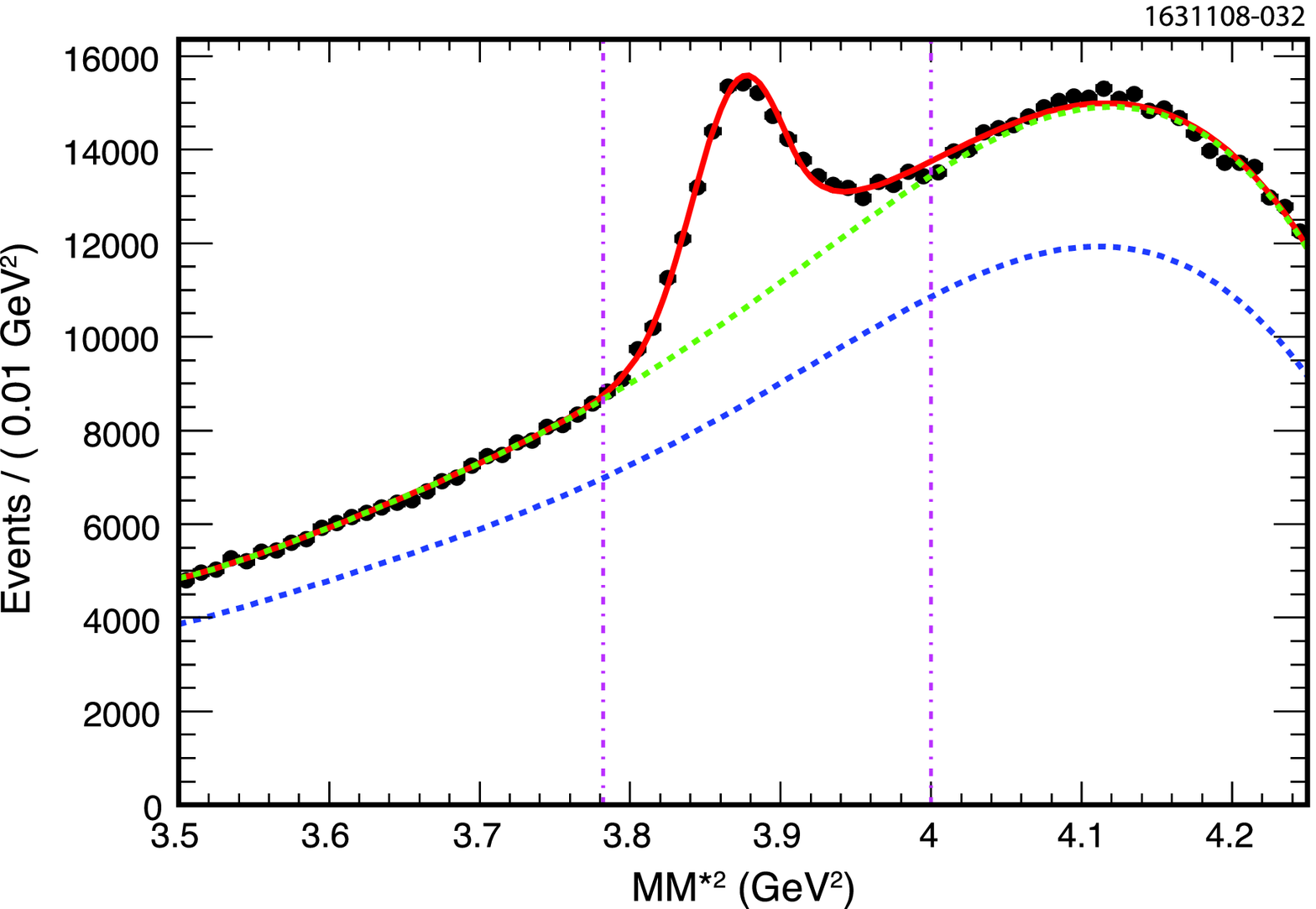}
\caption{The MM$^{*2}$ distribution summed over all modes. The
curves are fits to the number of signal events using the Crystal Ball function and two 5th order
Chebychev background functions (see text). The vertical lines show
the region of events selected for further analysis.}
\label{mms238-48_sig}
\end{figure}

We also determine the number of $D_s^-$ tags when the tag results
from the decay $D_s^{*-}\to \gamma D_s^-$. This procedure reduces
the background considerably in determining the tag yield and results
in counting half of the tags. Adopting this procedure we find a tag
yield that is higher by (1.2$\pm$2.3)\% from our nominal procedure,
consistent with the assigned systematic error.  The summed
MM$^{*2}$ distribution is shown in Fig.~\ref{mms238-48_sig}.

There is also a small enhancement of 5.2\% in our ability to find
tags in $\mu^+\nu$ (or $\tau^+\nu$, $\tau^+\to\pi^+\overline{\nu}$)
events (tag bias) as compared with events where the $D_s^+$ decays generically.
We determine this correction by using a Monte Carlo simulation for each tag mode
independently and then average the results based on the known tag fractions. We assign
a systematic error of
20\% giving a correction of $(5.2 \pm 1.0)$\%.

\subsection{Signal Reconstruction}
We next describe the search for $D_s^+\to\mu^+\nu$. We select events
within the MM$^{*2}$ region shown in Fig.~\ref{mms238-48_sig} for
further analysis. We note that the limits are rather wide. We use
this selection because the background in
the signal side is rather small and the errors are minimized by
taking as many tags as possible.

Candidate events
are selected that contain only a single extra track with opposite
charge to the tag. The track must make an angle
$>$25.8$^{\circ}$ with respect to the beam line to ensure that it is
well measured, and in addition we
require that there not be any neutral cluster detected in the
calorimeter, not associated with the tag,  with energy greater than 300 MeV (photon veto).
These cuts are highly
effective in reducing backgrounds. The photon energy cut is
especially useful to reject $D_s^+\to \pi^+\pi^0$, should this mode
be significant, and $D_s^+\to\eta\pi^+$.

Since we are searching for events where there is a single missing
neutrino, the missing mass squared, MM$^2$, evaluated by taking into
account the observed $\mu^+$, $D_s^-$, and $\gamma$ should peak at
zero; the MM$^2$ is computed as

\begin{equation}
\label{eq:mm2} {\rm MM}^2=\left(E_{\rm
CM}-E_{D_s}-E_{\gamma}-E_{\mu}\right)^2
           -\left({\bf p}_{\rm CM}-{\bf p}_{D_s}
           -{\bf p}_{\gamma}
           -{\bf p}_{\mu}\right)^2,
\end{equation}
where $E_{\mu}$ (${\bf p}_{\mu}$) are the energy
(momentum) of the candidate muon track and all other variables are the
same as defined in Eq.~(\ref{eq:mmss}).

We also make use of a set of kinematical constraints and fit each
event to two hypotheses one of which is that the $D_s^-$ tag is the
daughter of a $D_s^{*-}$ and the other that the $D_s^{*+}$ decays
into $\gamma D_s^+$, with the $D_s^+$ subsequently decaying into
$\mu^+\nu$. The kinematical constraints, in the $e^+e^-$ center-of-mass
frame, are
\begin{eqnarray}
\label{eq:constr}
&&{\bf p}_{D_s}+{\bf p}_{D_s^*}=0
\\\nonumber &&E_{\rm CM}=E_{D_s}+E_{D_s^*}\\\nonumber
&&E_{D_s^*}=\frac{E_{\rm
CM}}{2}+\frac{M_{D_s^*}^2-M_{D_s}^2}{2E_{\rm CM}}{\rm~or~}
E_{D_s}=\frac{E_{\rm CM}}{2}-\frac{M_{D_s^*}^2-M_{D_s}^2}{2E_{\rm
CM}}\\\nonumber &&M_{D_s^*}-M_{D_s}=143.8 {\rm ~MeV}.
\end{eqnarray}
In addition, we constrain the invariant mass of the $D_s^-$ tag to
the known $D_s$ mass. This gives us a total of 7 constraints. The
missing neutrino four-vector needs to be determined, so we are left
with a three-constraint fit. We perform a standard iterative fit
minimizing $\chi^2$. As we do not want to be subject to systematic
uncertainties that depend on understanding the absolute scale of the
errors, we do not make a $\chi^2$ cut but simply choose the photon
and the decay sequence in each event with the minimum $\chi^2$.

In this analysis, we consider three separate cases: (i) the track
deposits $<$~300 MeV in the calorimeter, characteristic of a
non-interacting pion or a muon; (ii) the track deposits $>$~300 MeV
in the calorimeter, characteristic of an interacting pion, and is
not consistent with being an electron; (iii) the track satisfies our
electron selection criteria; for a track to be called an electron, we require
that the momentum measurement in the tracking system and
the energy deposited in the CsI calorimeter are close to being equal within errors, and we also require that
$dE/dx$ and RICH information be consistent with expectations for an
electron. Then we separately study
the MM$^2$ distributions for these three cases. The separation
between muons and pions is not complete. Case (i) contains 98.8\% of
the muons but also 55\% of the pions, while case (ii) includes 1.2\%
of the muons and 45\% of the pions \cite{DptomunPRD}. Case (iii)
does not include any signal but is used later to search for
$D_s^+\to e^+\nu$. For cases (i) and (ii) we insist that the track
not be identified as a kaon.

\subsection{The Expected MM$^2$ Signal Spectrum}

\begin{figure}[hbt]
\centering
\includegraphics[width=5in]{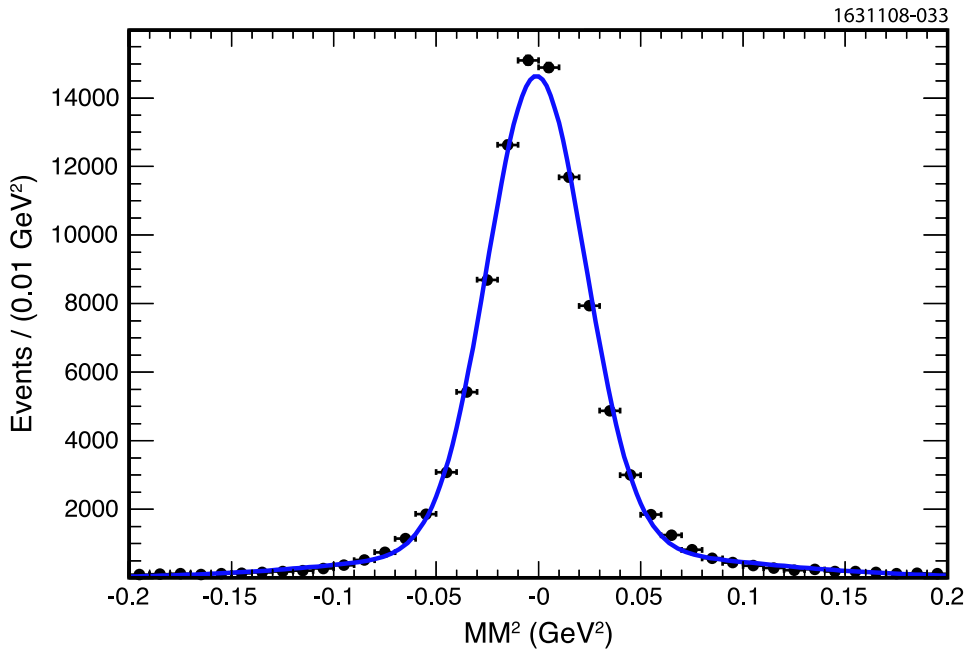}
\caption{The MM$^2$ resolution from Monte Carlo simulation for
$D_s^+\to\mu^+\nu$. The curve is the sum of two-Gaussians with means
constrained to be the same.} \label{mc-munu-res}
\end{figure}

For the $\mu^+\nu$ final state the MM$^2$ distribution can be
modeled as the sum of two Gaussians centered at zero (see
Eq.~(\ref{eq:twoGauss}). A Monte Carlo simulation of the MM$^2$ is
shown in Fig.~\ref{mc-munu-res}. A fit using the two-Gaussian shape
gives $\sigma_1=0.0240$ GeV$^2$, $\sigma_2=0.0851$ GeV$^2$,
$f=0.275$, which results in $\sigma=0.0346\pm 0.0002$ GeV$^2$.

\begin{figure}[hbt]
\centering
\includegraphics[width=4in]{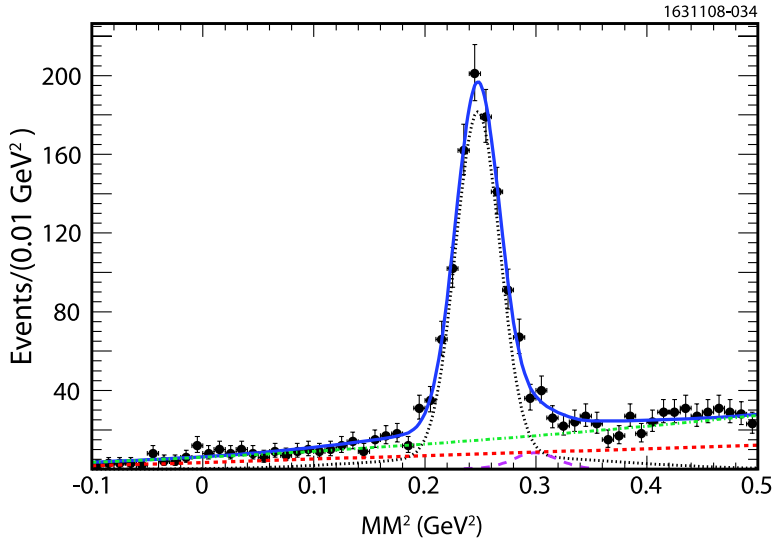}
\caption{The MM$^2$ distribution for events with an identified $K^+$
track. The kinematic fit has been applied. The data are shown as
points with error bars. The long-dashed curve shows the calculated yield of $\eta K^+$ events. The solid curve shows the results of a fit to the data, where the dotted curve is the
sum of two Gaussians centered at the square of the ${K}^0$ mass, and the
dashed and dot-dashed lines refer to the sideband, and combinatoric
backgrounds, respectively.} \label{Kmm2-data}
\end{figure}

We check the resolution using data. The mode $D_s^+\to
\overline{K}^0K^+$ provides an excellent testing ground.\footnote{In
this paper the notation $\overline{K}^0K^+$ refers to the sum of
$\overline{K}^0K^+$ and ${K^0}K^+$ final states.} We search for
events with at least one additional track identified as a kaon using
the RICH detector, in addition to a $D_s^-$ tag. We allow events
with no more than two other additional charged tracks, to allow for the
presence of $K^0$ decays, and we do not apply the greater than 300
MeV extra energy cut. The MM$^2$ distribution is shown in
Fig.~\ref{Kmm2-data}. Fitting this distribution to a two-Gaussian
signal shape gives a MM$^2$ resolution $\sigma=0.0338\pm0.0014$ GeV$^2$ in
agreement with Monte Carlo simulation which gives 0.0344$\pm$0.0003
GeV$^2$. (The backgrounds are discussed in section~\ref{sec:checks}.)
We note that the resolution here is larger than in our
previous work \cite{Previous}. This is mainly due to the use of
additional decay modes with photons, and enlarging of the solid
angle for candidate muons.

\begin{figure}[hbt]
\centering
\includegraphics[width=4in]{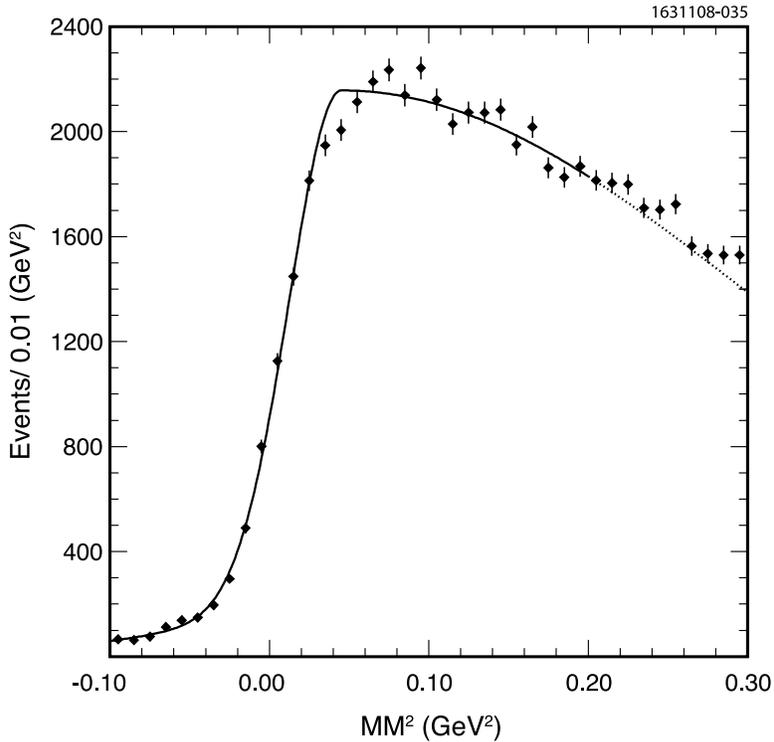}
\caption{The MM$^2$ distribution from Monte-Carlo simulation for
$D_s^+\to\tau^+\nu$, $\tau^+\to\pi^+\overline{\nu}$ at an $E_{\rm
CM}$ of 4170 MeV. The curve is a fit to the sum of two-Gaussians
with different widths on the low and high MM$^2$ sides.}
\label{mm2-taunu-pinu-mc}
\end{figure}

For the $\tau^+\nu$, $\tau^+\to\pi^+\overline{\nu}$ final state a
Monte Carlo simulation of the MM$^2$ spectrum is shown in
Fig.~\ref{mm2-taunu-pinu-mc}. The extra missing neutrino results in
a smeared distribution.

\subsection{MM$^2$ Spectra in Data}

\begin{figure}[htb]
 %\vskip 0.00cm
%\centerline{ \epsfxsize=3.0in
\centerline{ \epsfxsize=3.5in \epsffile{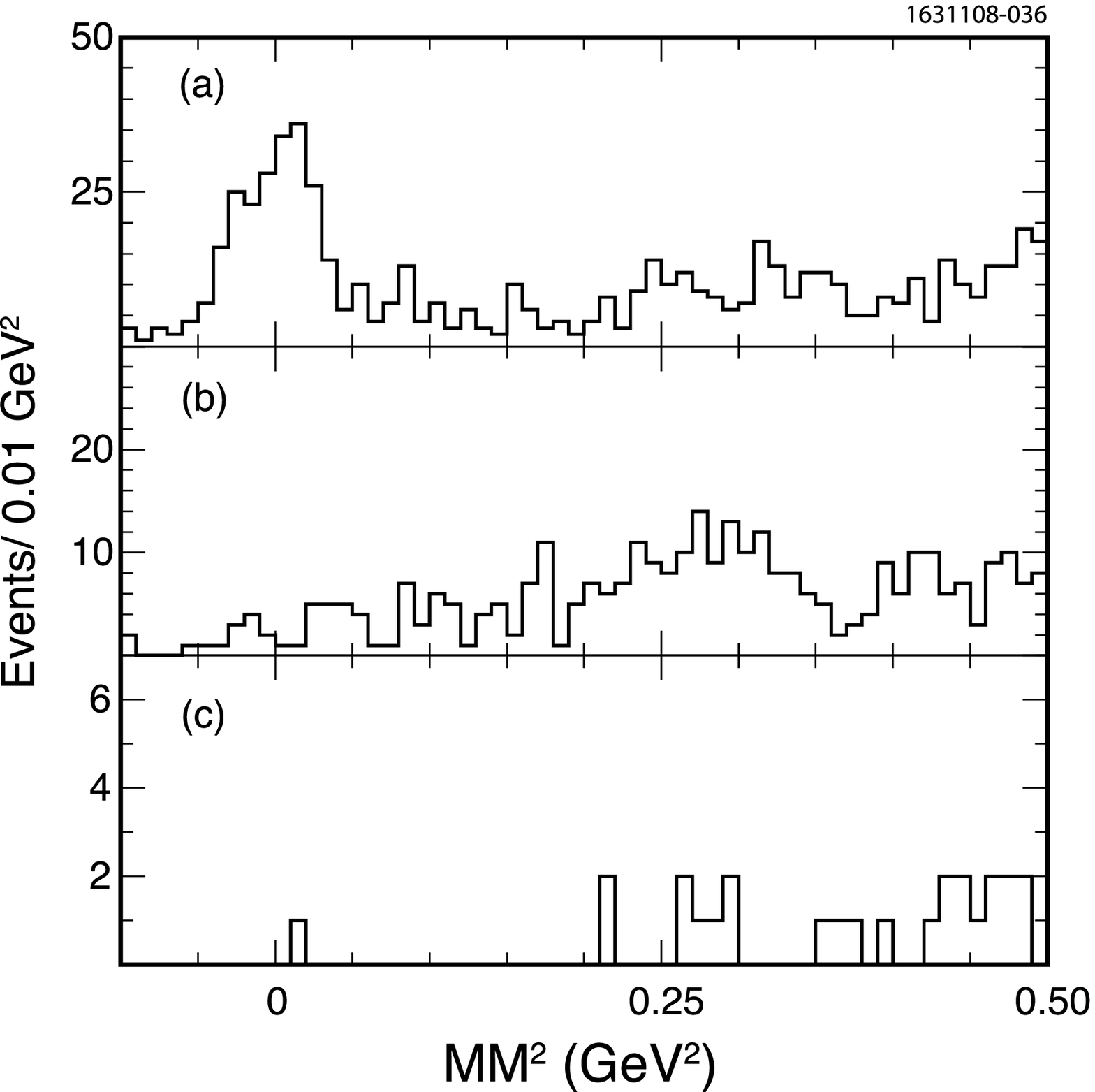}}
 \caption{The MM$^2$ distributions from data for events with a
$D_s^-$ reconstructed in a tag mode, an additional positively
charged track and no neutral energy clusters above 300 MeV.
(a) Case
(i) when the single track deposits $<$~300 MeV of energy in the
calorimeter. The peak near zero is from $D_s^+\to\mu^+\nu$ events.
(b) Case (ii): the track deposits $>$~300 MeV in the crystal calorimeter but
is not consistent with being an electron. (c) Case (iii): the track is
identified as an electron. } \label{mm2-data}
 \end{figure}
 The MM$^2$ distributions from data are shown in Fig.~\ref{mm2-data}.
The overall signal region we consider is $-0.1<$ MM$^2$ $<$ 0.20
GeV$^2$. The higher limit is imposed to exclude background from
$\eta\pi^+$ and ${K}^0\pi^+$ final states. There is a clear peak in
Fig.~\ref{mm2-data}(a) due to $D_s^+\to\mu^+\nu$. Furthermore, the
region between the $\mu^+\nu$ peak and 0.20 GeV$^2$ has events that
we will show are mainly due to the $D_s^+\to\tau^+\nu$, $\tau^+\to \pi^+\bar{\nu}$ decay.
The events in Fig.~\ref{mm2-data}(b) below 0.20 GeV$^2$ are also
mostly due to $\tau^+\nu$ decay.

\subsection{Background Evaluations}
\label{sec:background}

We consider the background arising from two sources: one from real
$D_s^+$ decays and the other from the background under the
single-tag signal peaks. For the latter, we obtain the background
from data by using a two-dimensional extended likelihood fit in
invariant mass and MM$^2$.

\begin{figure}[hbt]
\centering
\includegraphics[width=3in]{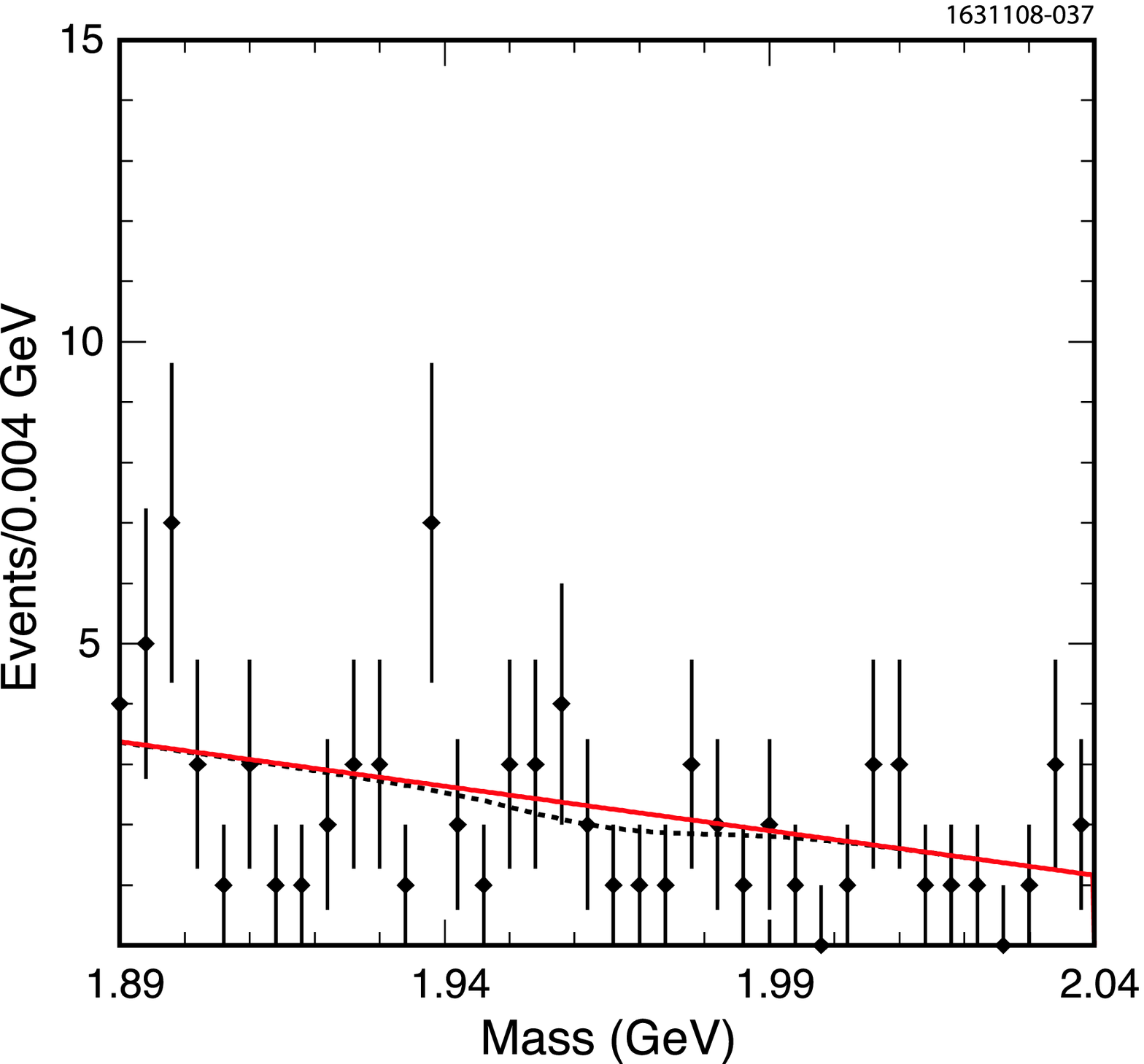}
\caption{The invariant $\pi^+\pi^0$ mass. The curves show results of
a fit using a linear background (solid) plus Gaussian signal
function (dashed), where the width of the Gaussian is fixed to a
value determined by Monte Carlo simulation. }
\label{pipi0-single-tag}
\end{figure}

The background from real $D_s^+$ decays is studied by identifying
each possible source mode by mode. For the $\mu^+\nu$ final state,
the only possible background within the signal region is
$D_s^+\to\pi^+\pi^0$.  (Recall that any such events are also heavily
suppressed by the extra photon energy veto of 300 MeV.) We search for
this mode by examining the  $\pi^+\pi^0$ invariant mass spectrum in
events where we have selected a $D_s^-$ tag based on invariant mass
only. See Fig.~\ref{pipi0-single-tag}. There is no peak at the
$D_s^+$ mass. Fitting to a linear background plus a Crystal-Ball
signal function, whose shape is fixed by Monte Carlo gives
-3.6$\pm$8.4 events. Setting the mean value to zero results
in an upper limit ${\cal {B}}(D_s^+\to\pi^+\pi^0) <3.8\times
10^{-4}$ at 90\% confidence level. Multiplying by our 43859 tags and
the 2\% inefficiency for detecting a $>$ 300 MeV photon from the
$\pi^0$ decays, results in an upper limit of 1/4 of an event in our
sample, which we ignore.

\begin{figure}[htbp]
 \vskip 0.00cm
\centerline{\epsfxsize=6.0in \epsffile{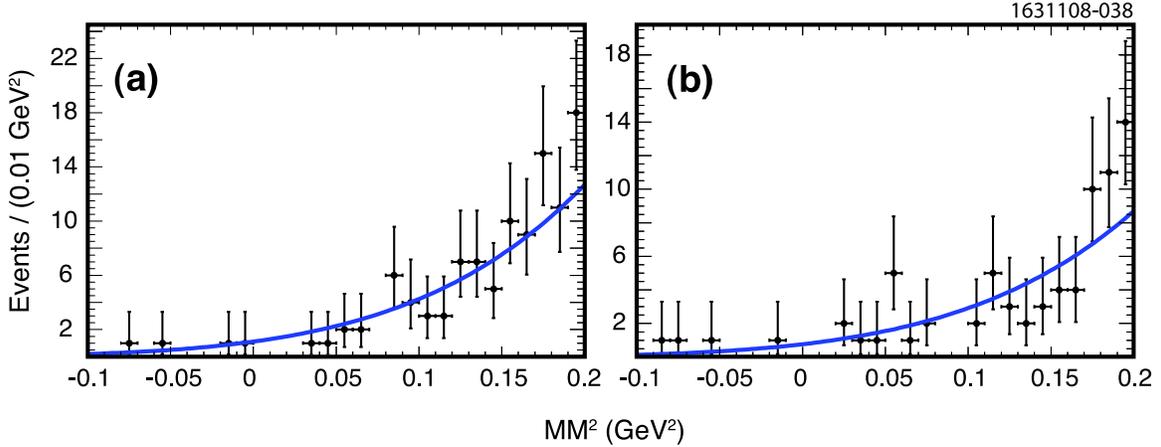}}
 \caption{The background rates for real $D_s^+$ decays from Monte Carlo as a function of MM$^2$ for
 case (i) and case (ii). The data are fit to quadratic polynomials.}
 \label{bgother}
 \end{figure}

For the $\tau^+\nu$, $\tau^+\to\pi^+\overline{\nu}$ final state the
real $D_s^+$ backgrounds include, in addition to the $\pi^+\pi^0$
background discussed above, semileptonic decays, possible
$\pi^+\pi^0\pi^0$ decays,  other $\tau^+$ decays, and small amounts
of $K^0\pi^+$ and $\eta\pi^+$ whose low MM$^2$ tails leak into the
signal region. Semileptonic
decays involving muons are equal to those involving electrons shown
in Fig.~\ref{mm2-data}(c). One event consistent with the electron
hypothesis is present. The $\pi^+\pi^0\pi^0$ background level is
estimated by considering the $\pi^+\pi^+\pi^-$ final state whose
measured branching fraction is (1.11$\pm$0.07$\pm$0.04)\% \cite{CLEO3pi}.
This mode has large contributions from $f_0(980)\pi^+$ and other
$\pi^+\pi^-$ resonant structures at higher mass \cite{focus-3pi}.
The $\pi^+\pi^0\pi^0$ mode will also have these contributions, but
the MM$^2$ opposite to the $\pi^+$ will be at large mass. The only
component that can potentially cause background is the non-resonant
component measured by FOCUS as (17$\pm$4)\% \cite{focus-3pi}. This
background has been evaluated by Monte Carlo simulation as have
backgrounds from other $\tau^+$ decays, $K^0\pi^+$ and $\eta\pi^+$.
 The backgrounds are
enumerated in Table~\ref{tab:back}. We show in Fig.~\ref{bgother}
the sum of all backgrounds and a fit to a quadratic polynomial over the MM$^2$
range of interest.

\begin{table}[htb]
\begin{center}
\caption{Background estimates for the data in the signal region
$-0.1<$MM$^2<0.2$ GeV$^2$. (We assume ${\cal{B}}(D_s^+\to\tau^+\nu)
= 6.2\pm0.7$\%.)\label{tab:back}}
\begin{tabular}{lccc}
 \hline\hline
    Final State  & $\cal{B}$(\%) & \# of events case(i) & \# of events case (ii)\\\hline
$\tau^+\to\pi^+\pi^0\overline{\nu}$ & 1.6$\pm$0.2 & 2.06$\pm$0.34 & 1.43$\pm$0.36 \\
$\tau^+\to\mu^+\nu\overline{\nu}$ & 1.1$\pm$0.1 & 1.60$\pm$0.24 & 0 \\
$D_s^+\to \pi^+\pi^0\pi^0$ & 1.1 (estimate) & 0.12 & 0.12 \\
$D_s^+\to K^0\pi^+$ & 0.24$\pm$0.03 & 1.3 $\pm$0.3 &1.1 $\pm$0.3\\
$D_s^+\to \eta\pi^+$ & 1.5 $\pm$ 0.2&1.1 $\pm$ 0.3 &0.9 $\pm$ 0.3\\
\hline
Sum & & 6.2$\pm$0.7 & 3.5$\pm$0.6\\
\hline\hline
\end{tabular}
\end{center}
\end{table}

\section{Leptonic Branching Fractions}

\begin{figure}[htbp]
 \vskip 0.00cm
\centerline{
\epsfxsize=6.0in \epsffile{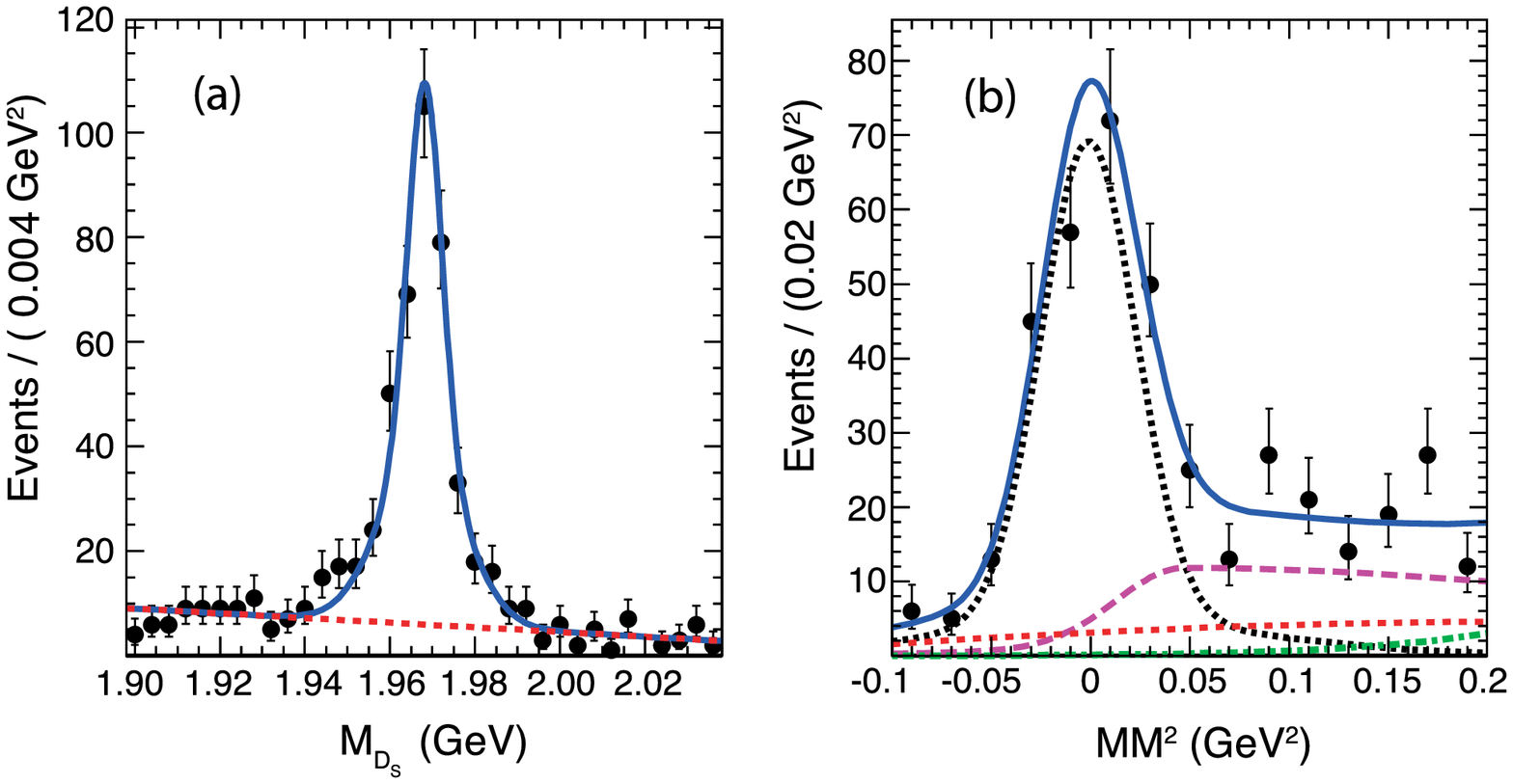}}
 \caption{The results of the two-dimensional fit to the  sum of case (i) and case (ii) data.
 The data are shown as points with error bars. (a) The projection of the invariant mass
 distribution; the straight dashed line shows the background while
  the curve is the sum of the background and a two-Gaussian signal function. (b) The projection
  of the MM$^2$ distribution; the dotted (black) curve is the two-Gaussian signal function
  for $\mu^+\nu$, the long-dashed (purple) curve shows the
  $\tau^+\nu$, $\tau^+\to\pi^+\overline{\nu}$ signal, the dashed (red) line shows the
  background from non-$D_s^-$ events below the signal peak, while the dot-dashed (green) curve shows the background from real $D_s^+$ events. The solid (blue) curve represents the sum of all contributions.}
 \label{try1-total}
 \end{figure}

The result of the two-dimensional unbinned maximum liklihood fit to the sum of the
MM$^2$ distributions for case (i) and case (ii)
is shown
in Fig.~\ref{try1-total}. The other dimension in the fit is the invariant mass spectrum. Here we constrain the $\tau^+\nu$/$\mu^+\nu$ ratio to the Standard
Model value.
From the fit we extract
235.5$\pm$13.8 $\mu^+\nu$ signal events. The efficiency is given by the
product of the tracking and particle identification efficiencies, equal to
86.7\%, the maximum extra photon energy cut of 300 MeV, equal to
98.7\% and the tag bias of 105.2\%, giving an overall efficiency of
90.0\%. Using our tag sample of 43859$\pm$936 events, we find an
effective branching ratio
\begin{equation}
{\cal B}^{\rm eff}(D_s^+\to \mu^+\nu)=(0.597\pm 0.037 \pm 0.017)\%
\label{eq:finalBR}
\end{equation}
The radiative correction of 1\% reduces this to
\begin{equation}
{\cal B}^{\rm eff}(D_s^+\to \mu^+\nu)=(0.591\pm 0.037 \pm 0.018)\%
\end{equation}
(From now on we will only quote radiatively corrected results in this paper.)
This is our most accurate result within the context of the Standard Model.

\begin{figure}[htbp]
 \vskip 0.00cm
\centerline{\epsfxsize=6.0in \epsffile{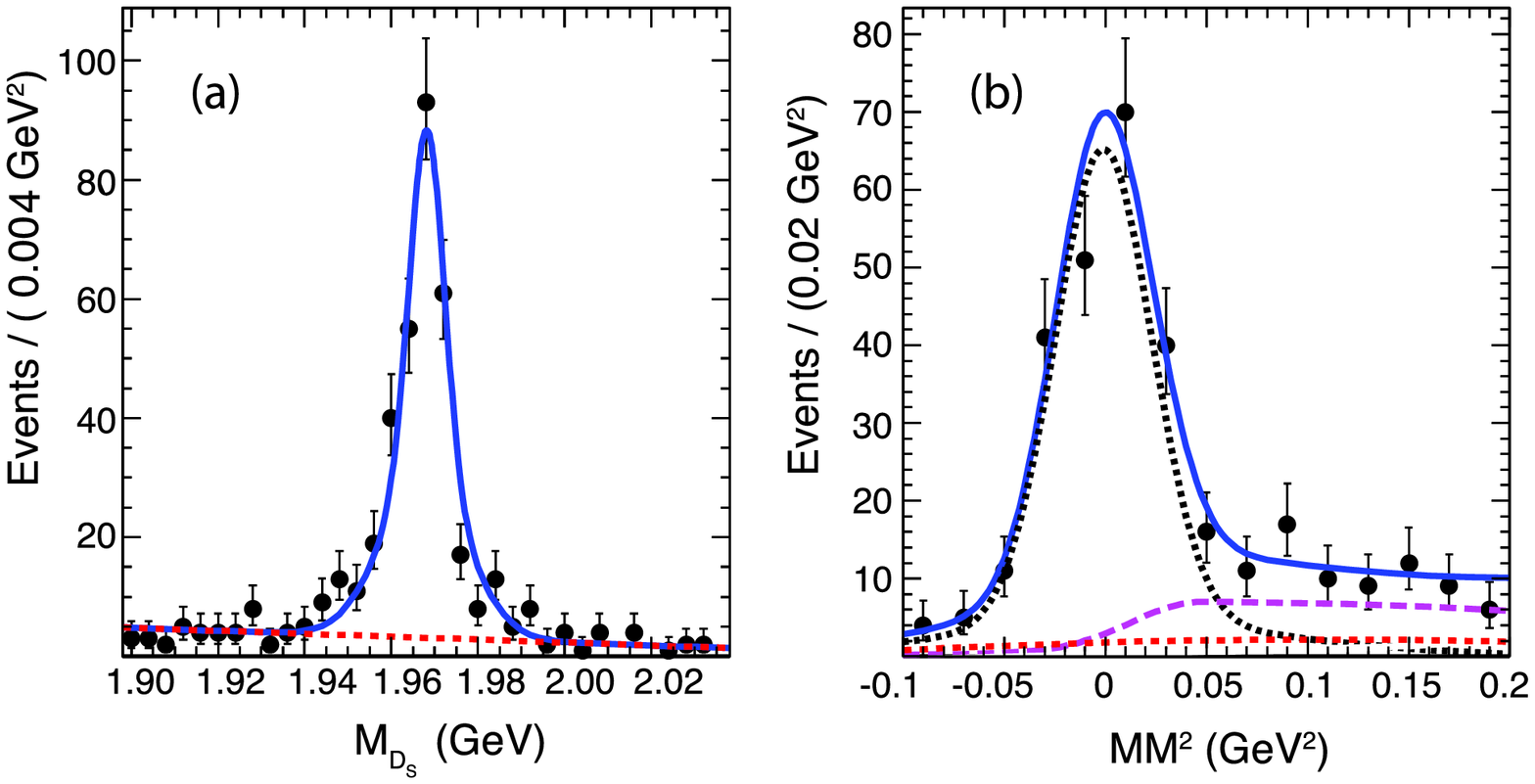}}
 \caption{The results of the two-dimensional fit to the case (i) data.
 The data are shown as points with error bars. (a) The projection of the invariant mass
 distribution; the straight dashed line shows the background while
  the curve is the sum of the background and a two-Gaussian signal function. (b) The projection
  of the MM$^2$ distribution; the dotted (black) curve is the two-Gaussian signal function
  for $\mu^+\nu$, the long-dashed (purple) curve shows the
  $\tau^+\nu$, $\tau^+\to\pi^+\overline{\nu}$ signal, the dashed (red) line shows the
  background from non-$D_s^-$ events below the signal peak, while the dot-dashed (green) curve shows the background from real $D_s^+$ events. The solid (blue) curve represents the sum of all contributions.}
 \label{mm2_case1}
 \end{figure}

We can also analyze the data by not constraining the $\tau^+\nu$/$\mu^+\nu$ ratio. We then fit the
case (i) distribution and find
\begin{equation}
{\cal B}(D_s^+\to \mu^+\nu)=(0.565\pm 0.045 \pm 0.017 )\%
\end{equation}
The results of the fit are shown in Fig.~\ref{mm2_case1}.

\begin{figure}[htbp]
 \vskip 0.00cm
\centerline{\epsfxsize=6.0in \epsffile{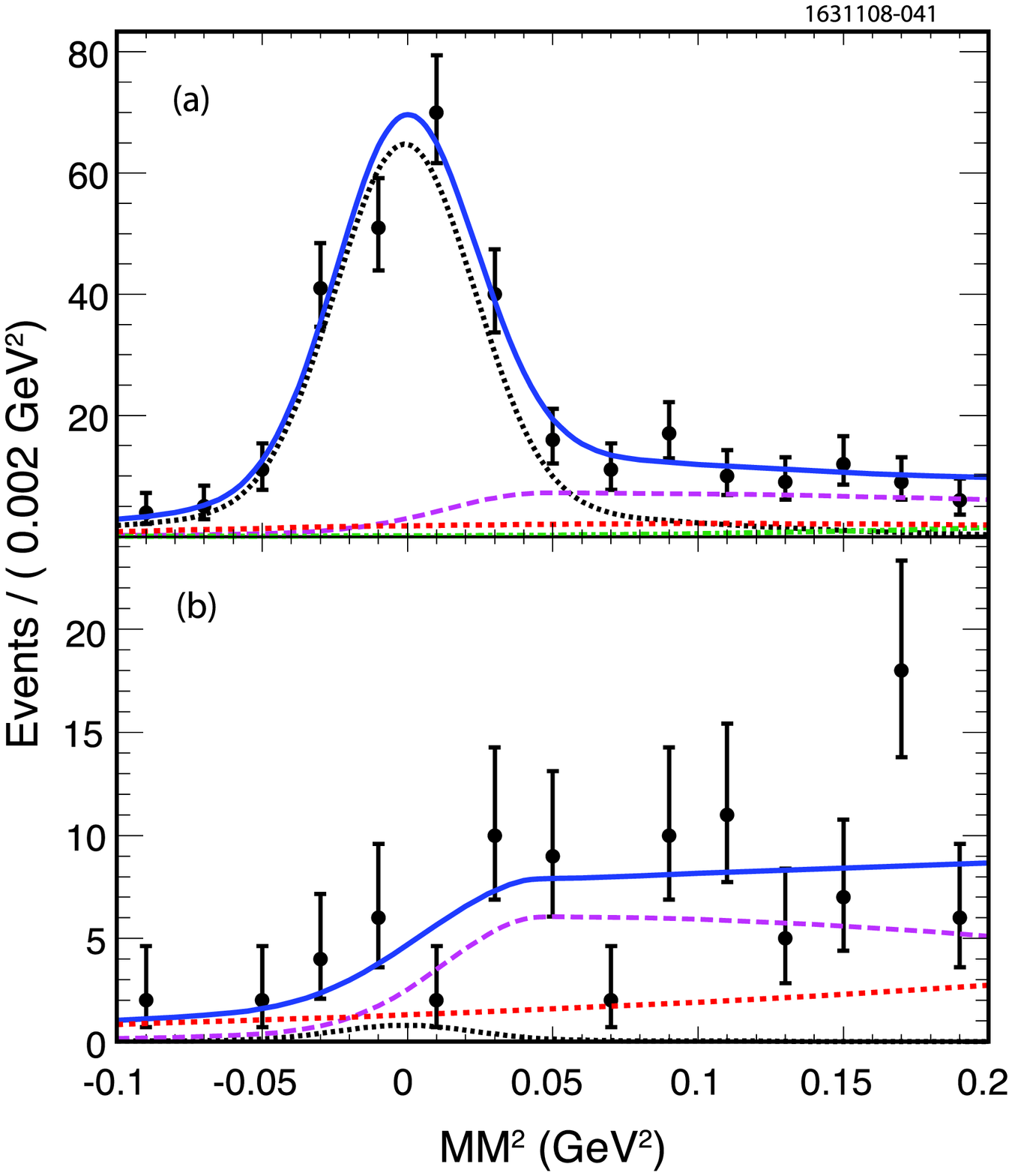}}
 \caption{The MM$^2$ distributions from the two-dimensional fit
 done simultaneously to (a) case (i) and (b) case (ii) data.
 The data are shown as points with error bars. The dotted (black) curve is the two-Gaussian signal function
  for $\mu^+\nu$, the long-dashed (purple) curve shows the
  $\tau^+\nu$, $\tau^+\pi^+\overline{\nu}$ signal, the dashed (red) line shows the
  background from non-$D_s^-$ events below the signal peak, while the dot-dashed (green) curve shows the background from real $D_s^+$ events. The solid (blue) curve represents the sum of all contributions.}
 \label{mm2_simultaneous}
 \end{figure}

By simultaneously fitting the case (i) and case (ii) distributions,
constraining the ratio of $\mu^+\nu$ events to be 98.8:1.2 and the
ratio of $\tau^+\nu$ events to be 55:45, in the two cases,
respectively, we find 125.6$\pm$15.7 $\tau^+\nu$,
$\tau^+\to\pi^+\overline{\nu}$ events. Using ${\cal{B}}(\tau^+\to
\pi^+\bar{\nu}$)=(10.90$\pm$0.07)\% \cite{PDG}, we measure
\begin{equation}
{\cal B}(D_s^+\to \tau^+\nu)=(6.42\pm 0.81 \pm 0.18)\%
\end{equation}
The fit results are shown in Fig.~\ref{mm2_simultaneous}.

The systematic errors on these branching fractions are given in
Table~\ref{tab:munusys}. The error on track finding is determined
from a detailed comparison of the simulation with double tag
events where one track is ignored. The particle identification on the
$\mu^+$ track arises from the fact that we veto kaons.
The error on the photon veto efficiency, due to the 300 MeV/c
extra shower energy cut, is determined using double tag events where
both $D_s^-$ and $D_s^+$ are reconstructed in the same modes that we
use for tagging. What we measure is a product of the efficiencies for
two events each with different tags. This is translated into an efficiency
for a single tags.
The error on the number of tags $\pm$2\% is assigned by varying the
fitting functions and ranges.
In addition there is a small error of
$\pm$0.6\% on the $\tau^+\nu$ branching fraction due to the
uncertainty on the $\tau^+$ decay fraction to
$\pi^+\overline{\nu}$. Additional systematic errors arising from
the background estimates are at the 1\% level. The error on
the radiative correction is taken as 100\% of its value of 1\%. When we use only one
of the two cases to find a result an additional 1\% error is included
due to the minimum ionization discrimination of 300 MeV in the calorimeter.
\begin{table}[htb]
\begin{center}
\caption{Systematic errors on determination of the $D_s^+\to
\mu^+\nu$ branching fraction. \label{tab:munusys}}
\begin{tabular}{lc} \hline\hline
   Error Source & Size (\%) \\ \hline
Track finding &0.7 \\
Particle identification of $\mu^+$ & 1.0\\
MM$^2$ width & 0.2 \\
Photon veto & 0.4 \\
Background & 1.0\\
Number of tags& 2.0\\
Tag bias & 1.0\\
Radiative Correction & 1.0\\
\hline
Total & 3.0\\
 \hline\hline
\end{tabular}
\end{center}
\end{table}

Lepton universality in the Standard Model requires that the ratio
$R$ from Eq.~(\ref{eq:tntomu}) be equal to a value of 9.76. We measure
\begin{equation}
R\equiv \frac{\Gamma(D_s^+\to \tau^+\nu)}{\Gamma(D_s^+\to
\mu^+\nu)}= 11.4\pm 1.7 \pm 0.2~. \label{eq:tntomu2}
\end{equation}
Here the systematic error is dominated by the uncertainty on the
minimum ionization cut that we use to separate the $\mu^+\nu$ and
$\tau^+\nu$ regions at 300 MeV. We take this error as 2\%, since a
change here affects both the numerator and denominator. The ratio is
consistent with the Standard Model prediction. Current results on
$D^+$ leptonic decays also show no deviations \cite{ournewDp}.

We also measure the CP violating asymmetry. The fit constraining the
$\tau^+\nu/\mu^+\nu$ ratio to the SM value of 9.76 yields 124.5$\pm$9.9
$\mu^+\nu$ and 110.8$\pm$9.6 $\mu^-\bar{\nu}$ events. We also find 21807$\pm$581
$D_s^-$ tags and 21370$\pm$581 $D_s^+$ tags. Then
\begin{equation}
\frac{\Gamma\left(D_s^+\to\mu^+\nu\right)-\Gamma\left(D_s^-\to\mu^-\bar{\nu}\right)}
{\Gamma\left(D_s^+\to\mu^+\nu\right)+\Gamma\left(D_s^-\to\mu^-\bar{\nu}\right)} = 0.048\pm0.061~,
\end{equation}
showing no evidence of CP violation.

 The
one detected electron opposite to our tags allows us to set an
upper limit of
\begin{equation}
{\cal{B}}(D_s^+\to e^+\nu)< 1.2\times 10^{-4}
\end{equation}
at 90\% confidence level; this is also consistent with Standard
Model predictions and lepton universality.

\section{Checks of the Method}
\label{sec:checks}

We perform an overall check of our procedures by measuring
${\cal{B}}(D_s^+\to \overline{K}^0K^+)$ which has been previously
determined. For this measurement we compute the MM$^2$
[Eq.~(\ref{eq:mm2})] using events with an additional charged track but
here identified as a kaon. These track candidates have momenta of
approximately 1 GeV/c; here our RICH detector has a pion to kaon
fake rate of 1.1\% with a kaon detection efficiency of 88.5\%
\cite{RICH}. For this study, we do not veto events with neutral
energy deposits $>$300 MeV, or with less than three additional
tracks beyond the tag, because of the presence of the ${K^0}$.

Events from the $\eta\pi^+$ mode where the $\pi^+$ fakes a
$K^+$ are very rare and would not peak at the proper MM$^2$. The mode
$\eta K^+$ does contribute a background at a somewhat higher  MM$^2$ of 0.30 GeV$^2$
and causes a small asymmetric tail on the high side of the peak. The branching fraction measured by CLEO is (0.14$\pm$0.03)\% \cite{CLEO-etaK}. We predict a total of 47$\pm$10 events from this source, that we include as a fixed component in our fit.

We perform the same two-dimension fit in invariant mass and MM$^2$
as used for the $\mu^+\nu$ signal, except that here the MM$^2$
is calculated with respect to the $K^+$ hypothesis, and an additional $\eta K^+$ component is added. The MM$^2$ distribution
for events in the signal MM$^{*2}$ region is shown in Fig.~\ref{Kmm2-data}. The
peak near 0.25 GeV$^2$ is due to the decay mode of interest. The backgrounds
are the same as defined for the $\mu^+\nu$ distributions above.

The fit yields 1036$\pm$41
events. In order to
compute the branching fraction we use the efficiency of detecting
the kaon track, 77.0\%, including radiation \cite{gammamunu}, the
particle identification efficiency of 88.5\%, and take into account
that it is easier to detect tags in events containing a
$\overline{K}^0K^+$ decay than in the average $D_sD_s^*$ event due
to the track and photon multiplicities, which gives a 3\%
correction.\footnote{The tag bias is less here than in the
$\mu^+\nu$ case because of the $K^0$ decays and interactions in the
detector.} These rates are estimated by using Monte Carlo
simulation. We determine
\begin{equation}
\label{eq:KK0}
 {\cal{B}}(D_s^+\to \overline{K}^0K^+)=(3.06\pm0.14\pm 0.09)\%,
\end{equation}
where the systematic errors are listed in Table~\ref{tab:sysKK}.
We estimate the error from the signal shape by taking the change
in the number of events when varying the signal width of the
two-Gaussian function by $\pm 1\sigma$. The error on the
background shapes is given by varying the shape of the background
fit. The error on the particle identification efficiency is
measured using two-body $D^0$ decays \cite{RICH}. The other errors
are the same as described in Table~\ref{tab:munusys}. Again, the
largest component of the systematic error arises from the number
of tag events (2\%). In fact, to use this result as a check on our
procedures, we need only consider the systematic errors that are
different here than in the $\mu^+\nu$ case. Those are due only to
the signal and background shapes, the $\eta K^+$ contribution and the particle identification
cut. Those systematic errors are small.

To determine absolute branching fractions of charm mesons, CLEO-c
uses a method where both particles are fully reconstructed (so
called ``double tags") and the rates are normalized using events
where only one particle is fully reconstructed. Our
result using this method for ${\cal{B}}(D_s^+\to K_S^0K^+)$=(1.49$\pm
0.07\pm 0.05)$\%, which when doubled becomes (2.98$\pm 0.14\pm
0.10)$\% \cite{CLEO3pi}. This is in excellent agreement with the
number in Eq.~(\ref{eq:KK0}).
 These results are not independent.

\begin{table}[htb]
\begin{center}

\caption{Systematic errors on determination of the $D_s^+\to
\overline{K}^0K^+$ branching fraction. \label{tab:sysKK}}
\begin{tabular}{lc} \hline\hline
   Error Source & Size (\%) \\ \hline
Track finding &0.7 \\
Particle identification of $\mu^+$ & 1.0\\
$\eta K^+$ branching fraction & 0.6\\
MM$^2$ width & 0.2 \\
Background & 1.0\\
Number of tags& 2.0\\
Tag bias& 1.0\\\hline
Total & 2.8\\
 \hline\hline
\end{tabular}
\end{center}
\end{table}

We also performed the entire analysis on a Monte Carlo sample that
corresponds to an integrated luminosity 8 times larger than the data
sample. The input branching fraction in the Monte Carlo is 0.61\%
for $\mu^+\nu$ and 5.99\% for $\tau^+\nu$, while our analysis
measured ${\cal{B}}^{\rm
eff}(D_s^+\to\mu^+\nu)=$(0.607$\pm$0.013)\% for $\mu^+\nu$ and
$\tau^+\nu$ combined. The individual rates are
 ${\cal{B}}(D_s^+\to\mu^+\nu)=$(0.615$\pm$0.016)\%,
and ${\cal{B}}(D_s^+\to\tau^+\nu)=$(6.02$\pm$0.27)\%.

\section{The Decay Constant and Conclusions}

 Using our most precise value for ${\cal{B}}(D_s^+\to \mu^+\nu)$
 from Eq.~(\ref{eq:finalBR}), that is derived using both our $\mu^+\nu$ and $\tau^+\nu$
 samples,
 and Eq.~(\ref{eq:equ_rate}) with a $D_s$ lifetime of (500$\pm$7)$\times 10^{-15}\,{\rm s}$ \cite{PDG}, we extract
 \begin{equation}
 f_{D_s^+}=263.3\pm 8.2\pm 3.9{~\rm MeV}.
 \end{equation}
This result has been radiatively corrected. These results supersede
all our previous measurements of the $D_s^+\to\mu^+\nu$ and
$D_s^+\to\tau^+\nu$, $\tau^+\to\pi^+\nu$ branching fractions which
use data samples that are subsumed in this paper. Using the CLEO-c result
based on an analysis of $D_s^+\to\tau^+\nu$, $\tau^+\to e^+\nu\bar{\nu}$ \cite{CLEO-CSP},
of $f_{D_s^+}=(252.5\pm 11.1\pm 5.2)$ MeV, we derive a CLEO-c average value of
\begin{equation}
 f_{D_s^+}=259.5\pm 6.6\pm 3.1{~\rm MeV}.
 \end{equation}

We combine with our $D^+$ result \cite{ournewDp}
\begin{equation}
f_{D^+}=205.8\pm 8.5\pm 2.5{\rm ~MeV}
\end{equation}
 and find a  value for
\begin{equation}
\displaystyle{\frac{f_{D_s^+}}{f_{D^+}}=1.26\pm 0.06\pm 0.02},
\end{equation}
where only a small part of the systematic error cancels in the ratio
of our two measurements. Our new measurements are compared with
other measurements in Table~\ref{tab:fDs}.

\begin{table}[htb]
\begin{center}

\caption{Our results for ${\cal{B}}(D_s^+\to \mu^+\nu)$,
${\cal{B}}(D_s^+\to \tau^+\nu)$, and $f_{D_s^+}$ compared with
previous measurements. Results have been updated for the new value
of the $D_s$ lifetime of 0.5 ps \cite{PDG}. ALEPH combines both
measurements to derive a value for the decay
constant. (This table adopted from Table I of ref.~\cite{Rosner-Stone}.) \label{tab:fDs}}
\begin{tabular}{llccc}\hline\hline
Exp. & Mode  & ${\cal{B}}$& ${\cal{B}}_{\phi\pi}$ (\%) & $f_{D_s^+}$ (MeV) \\
\hline
CLEO-c & $\mu^+\nu$ & $(5.65\pm 0.45\pm 0.17)\cdot 10^{-3}$ & & $257.3\pm 10.3\pm 3.9$\\
CLEO-c & $\tau^+\nu$ & $(6.42\pm 0.81\pm 0.18)\cdot 10^{-2}$&& $278.7\pm 17.1 \pm 3.8 $ \\
CLEO-c & \multicolumn{2}{c}{combined above 2 results using SM} & & $263.3\pm 8.2\pm 3.9$  \\
CLEO-c  \cite{CLEO-CSP}& $\tau^+\nu$ & $(5.30\pm 0.47\pm 0.22)\cdot 10^{-2}$&& $252.5\pm 11.1 \pm 5.2$ \\
CLEO-c & \multicolumn{2}{c}{combined all CLEO-c results}  & &$259.5\pm 6.6\pm 3.1$  \\
Belle$^{\dagger}$ \cite{Belle-munu}  & $\mu^+\nu$ & $(6.38\pm 0.76\pm 0.52)\cdot 10^{-3}$&& $274\pm 16 \pm 12 $ \\
\hline
\multicolumn{3}{l}{Average of CLEO and Belle results above, radiatively corrected} & & $261.2\pm 6.9$\\
 \hline
CLEO \cite{CLEO}& $\mu^+\nu$ &$(6.2\pm 0.8\pm 1.3 \pm 1.6)\cdot
10^{-3}$&
3.6$\pm$0.9&$273\pm19\pm27\pm33$\\
BEATRICE \cite{BEAT} & $\mu^+\nu$ &$(8.3\pm 2.3\pm 0.6 \pm
2.1)\cdot 10^{-3}$& 3.6$\pm$0.9&$312\pm43\pm12 \pm39$\\
ALEPH \cite{ALEPH}& $\mu^+\nu$ &$(6.8\pm 1.1\pm 1.8)\cdot 10^{-3}$ & 3.6$\pm$0.9& $282\pm 19\pm 40$ \\
ALEPH \cite{ALEPH}& $\tau^+\nu$ &$(5.8\pm 0.8\pm 1.8)\cdot 10^{-2}$ & &  \\
L3 \cite{L3} &$\tau^+\nu$ & $(7.4\pm 2.8 \pm 1.6\pm 1.8)\cdot 10^{-2}$ & & $299\pm 57\pm 32 \pm 37$  \\
OPAL \cite{OPAL} & $\tau^+\nu$ & $(7.0\pm 2.1 \pm 2.0)\cdot 10^{-2}$ & & $283\pm 44\pm 41$  \\
BaBar \cite{Babar-munu} & $\mu^+\nu$& $(6.74\pm 0.83\pm 0.26 \pm
0.66)\cdot 10^{-3}$ & 4.71$\pm$0.46 &
 $283\pm 17 \pm 7 \pm 14$\\\hline\hline
\end{tabular}
\end{center}
$\dagger$ This result has been radiatively corrected by multiplying
the measured branching ratio by 99\%.

\end{table}

 Most measurements
of $D_s^+\to\ell^+\nu$ are normalized with respect to ${\cal{B}}
(D_s^+\to\phi\pi^+)\equiv {\cal{B}}_{\phi\pi}$.
 An exception is the OPAL measurement which is
normalized to the $D_s$ fraction in $Z^0$ events that is derived
from an overall fit to heavy flavor data at LEP \cite{HFAG}. It
still, however, relies on absolute branching fractions that are
hidden by this procedure, and the estimated error on the
normalization is somewhat smaller than that indicated by the error
on ${\cal{B}}_{\phi\pi}$ available at the time of their
publication. The L3 measurement is normalized taking the fraction
of $D_s$ mesons produced in $c$ quark fragmentation as
0.11$\pm$0.02, and the ratio of $D_s^*/D_s$ production of
0.65$\pm$0.10. The ALEPH results use ${\cal{B}}_{\phi\pi}$ for
their $\mu^+\nu$ results and a similar procedure as OPAL for their
$\tau^+\nu$ results. We note that the recent BaBar result uses a
larger ${\cal{B}}_{\phi\pi}$ than the other results.
The CLEO-c determination of $f_{D_s^+}$ using the modes in this paper is the most accurate to
date.

Theoretical models that predict $f_{D_s^+}$ and the ratio
$\frac{f_{D_s^+}}{f_{D^+}}$ are listed in Table~\ref{tab:Models}.
\begin{table}[htb]
\begin{center}

\caption{Theoretical predictions of $f_{D^+_s}$, $f_{D^+}$, and
$f_{D_s^+}/f_{D^+}$. QL indicates quenched lattice calculations. (This table
adopted from Table II of ref.~\cite{Rosner-Stone}.)}
\label{tab:Models}
\begin{tabular}{lccl} \hline\hline
    Model &$f_{D_s^+}$ (MeV) &  $f_{D^+}$ (MeV)          &  ~~~~~$f_{D_s^+}/f_{D^+}$           \\\hline
Lattice (HPQCD+UKQCD) \cite{Lat:Foll} & $241\pm 3$ & $208\pm 4$ &
$1.162\pm 0.009$\\
 Lattice (FNAL+MILC+HPQCD)  \cite{Lat:Milc} &
 $249 \pm 3 \pm 16 $&$201\pm 3 \pm 17 $&$1.24\pm 0.01\pm 0.07$ \\
%Lattice ($n_f$=2) (CP-PACS) \cite{Lat:CPPACS} &
%$202\pm 12^{+20}_{-25}$& 1.18$\pm$0.09$\pm$xx \\
QL (QCDSF) \cite{QCDSF} &
$220\pm 6 \pm 5 \pm 11$ &$206 \pm 6\pm 3\pm 22 $&$1.07\pm 0.02\pm 0.02$ \\
 QL (Taiwan) \cite{Lat:Taiwan} &
$266\pm 10 \pm 18$ &$235 \pm 8\pm 14 $&$1.13\pm 0.03\pm 0.05$ \\
QL (UKQCD) \cite{Lat:UKQCD}&$236\pm 8^{+17}_{-14}$ & $210\pm 10^{+17}_{-16}$ & $1.13\pm 0.02^{+0.04}_{-0.02}$\\
QL \cite{Lat:Damir} & $231\pm 12^{+6}_{-1}$&$211\pm 14^{+2}_{-12}$ &
$1.10\pm 0.02$\\
QCD Sum Rules \cite{Bordes} & $205\pm 22$ & $177\pm 21$ & $1.16\pm
0.01\pm 0.03$\\
QCD Sum Rules \cite{Chiral} & $235\pm 24$&$203\pm 20$ & $1.15\pm 0.04$ \\
Field Correlators \cite{Field} & $210\pm 10$&$260\pm 10$ & $1.24\pm 0.03$ \\
Quark Model \cite{Quarkmodel}&268 &$234$  & 1.15 \\
Quark Model \cite{QMII}&248$\pm$27 &$230\pm$25  & 1.08$\pm$0.01 \\
LFQM (Linear) \cite{Choi} & 211 & 248 & 1.18\\
LFQM (HO) \cite{Choi} &194 & 233  & 1.20\\
LF-QCD \cite{LF-QCD} & 253& 241  & 1.05 \\
Potential Model \cite{Equations} & 241& 238  & 1.01 \\
Isospin Splittings \cite{Isospin} & & $262\pm 29$ & \\
\hline\hline
\end{tabular}
\end{center}
\end{table}
Upper limits on $f_{D^+}$ and $f_{D_s}$ of 230 and 270 MeV, respectively,
have been determined using two-point correlation functions by Khodjamirian \cite{Kho}.
Our result for $f_{D_s}$ is higher than most theoretical
expectations. The average of this new $f_{D_s^+}$ result with the
CLEO-c result based on the decay mode $D_s^+\to\tau^+\nu$,
$\tau^+\to e^+\nu\overline{\nu}$,
is ($259.5\pm 7.3$) MeV (see Table~\ref{tab:fDs}). This
rate differs by 2.3 standard deviations from the Follana \etal.
prediction \cite{Lat:Foll}, and is close to, but does not saturate the Khodjamirian bound.
(Averaging in the Belle result for
$D_s^+\to\mu^+\nu$, which is also an absolute branching fraction
measurement raises the difference to 2.6 standard deviations.)
If the difference with Follana \etal ~were to
persist with the advent of more precise measurements it could
be explained by a deficiency in the Lattice calculation, or physics beyond the Standard
Model. (We note that both unquenched
LQCD calculations agree with CLEO's result for $f_{D^+}$.) Either of
the two leptoquark models of Dobrescu and Kronfeld could explain a
discrepancy in the $D_s^+$ case \cite{Dobrescu-Kronfeld}. They also
have a charged Higgs model. Hewett \cite{Hewett}, and Akeroyd and Chen \cite{Akeroyd}
pointed out that leptonic decay widths are modified by new physics.
Specifically, for the $D^+$ and $D^+_s$, in the case of the
two-Higgs doublet model (2HDM), Eq.~(\ref{eq:equ_rate}) is modified
by a factor $r_q$ multiplying the right-hand side:

%$$
%r_q=\left[1-M^2_{D_q}\left({{\tan\beta}\over
%m_{H^+}}\right)^2\left({m_q \over {m_c+m_q}}\right)\right]^2,
%\EQN{}
%$$

\begin{equation}
r_q=\left[1+\left(1\over{m_c+m_q}\right)\left({M_{D_q}\over
M_{H^+}}\right)^2 \left(m_c-m_q\tan^2\beta\right)\right]^2,
\end{equation}
where $m_{H^+}$ is the charged Higgs mass, $M_{D_q}$ is the mass of
the $D$ meson (containing the light quark $q$), $m_c$ is the charm
quark mass, $m_q$ is the light-quark mass, and $\tan\beta$ is the
ratio of the vacuum expectation values of the two Higgs doublets.
(Here we modified the original formula of \cite{Akeroyd} to take into account the
charm quark coupling \cite{KronPC}.) To get an enhancement in the
rate, the $m_c$ term must be inducing the effect, which implies that
both the $D^+$ and $D_s^+$ would see a similar effect in
contradiction to the trends in our data. Another explanation in a
model based on R-parity violating supersymmetry has been given by
Kundu and Nandi \cite{Kundu-Nandi}.

\section{Acknowledgments}

We gratefully acknowledge the effort of the CESR staff
in providing us with excellent luminosity and running conditions.
D.~Cronin-Hennessy and A.~Ryd thank the A.P.~Sloan Foundation.
This work was supported by the National Science Foundation,
the U.S. Department of Energy,
the Natural Sciences and Engineering Research Council of Canada, and
the U.K. Science and Technology Facilities Council.
We thank C. Davies, B. Dobrescu, A. Kronfeld, P. Lepage,  P. Mackenzie,
R. Van de Water, and R. Zwicky for useful discussions.

\afterpage{\clearpage}

\end{document}